\documentclass[pra,twocolumn,superscriptaddress,floatfix,amsmath,nofootinbib,amssymb]{revtex4-1}

\usepackage{graphicx}
\usepackage{bm}
\usepackage{verbatim}
\usepackage{mathrsfs}
\usepackage{color}
\usepackage{paralist}
\usepackage{sidecap}
\usepackage{wrapfig}

\newcommand{\ii}{\mathrm{i}}

\newcommand{\ket}[1]{| {#1} \rangle}

\newcommand{\bra}[1]{\left\langle {#1} \right|}
	
\newcommand{\braket}[2]{\left\langle {#1}\left|{#2}\right.\right\rangle}

\begin{document}

\title{Mode invisibility as a quantum non-demolition measurement of coherent light}
\author{Marvellous Onuma-Kalu}
\email{monumaka@uwaterloo.ca}
\affiliation{Department of Physics \& Astronomy, University of Waterloo,  Ontario Canada N2L 3G1}
\author{Robert B. Mann}
\email{rbmann@uwaterloo.ca}
\affiliation{Department of Physics \& Astronomy, University of Waterloo,  Ontario Canada N2L 3G1}
\author{Eduardo Mart\'{i}n-Mart\'{i}nez}
\email{emartinm@uwaterloo.ca}
\affiliation{Institute for Quantum Computing, University of Waterloo, Waterloo, Ontario, N2L 3G1, Canada}
\affiliation{Department of Applied Math, University of Waterloo, Waterloo, Ontario, N2L 3G1, Canada}
\affiliation{Perimeter Institute for Theoretical Physics, 31 Caroline St N, Waterloo, Ontario, N2L 2Y5, Canada}

\begin{abstract}
We exploit geometric properties of quantum states of light in optical cavities to carry out quantum non demolition measurements. We generalize the ``mode-invisibility'' method to obtain information about the Wigner function of a squeezed coherent state in a non-destructive way. We also simplify the application of this non demolition technique to measure single-photon and few-photon states.
\end{abstract}

\maketitle

\section{Introduction}
Quantum nondemolition (QND) measurements \cite{Braginski,firstdemo,secondemo} try to obtain the maximum information possible about a quantum system while minimizing the measurement back action. Since its conceptual introduction, there has been a lot of theoretical \cite{PPRA} and experimental \cite{nature1,secondemo,nature3,singlephotons,revmods} progress in this field. QND measurements of the number of photons in Fock states of light  have been very successfully implemented in quantum optical settings \cite{serge}. 

In our earlier work \cite{marvy2013} we proposed a quantum non demolition measurement of Fock states of light taking advantage of the spatial symmetry of the modes of the field in an optical cavity. We called this technique ``mode invisibility''.  Using this technique we showed that an atomic probe, on resonance with the target field mode we want to measure, can be sent through a cavity in such a way that the state of light in that mode is not altered, but at the same time the atom acquires a non-negligible phase easily appreciable in an atomic interferometry experiment. We exploited the mode-invisibility  technique to suggest that a setting of two optical cavities---one containing a known state of light and another one containing the state of light that we want to probe---allows for the effective distinction of Fock states containing very few photons.

Unfortunately Fock states of light are difficult to generate.  We are therefore led to consider using coherent states that are experimentally much more controllable and easier to realize. In this article we  extend the mode-invisibility technique to the non-demolition measurement of coherent states of light  \cite{coherentstate, coinco} and squeezed coherent states \cite{squeezedlight}. We will show that it is indeed possible to obtain information about an unknown Fock state without having  access to another Fock state.  

This advantage extends to other states of light.  We will demonstrate ways in which  the mode-invisibility technique can yield information (other than the average photon content) about more complex states of light, such as squeezed coherent states. In general it is not obvious that the adiabatic approximation can be satisfied in the course of obtaining information about the physical parameters of  squeezed coherent states. However, we will see that the adiabatic approximation is indeed fulfilled.

What is more, we will discuss the possibility of gaining information (in a non-destructive way) about some features of the Wigner function of more general states of light, such as the relative difference in the phase of a squeezing and a phase space displacement.  


\section{Criteria for QND measurement}
We remark here that the key features of a QND measurement include its ability to preserve useful information for subsequent processing and  its repeatability, in which quantum-state evolution into a different state is prohibited and successive measurement yields the same result as the first measurement.

In a general  measurement scheme, the system to be measured is coupled to a probe system, and the interaction of the two systems correlates the states of the probe and measured system. For the measurement to be a QND measurement, the measurement scheme should satisfy a set of conditions  \cite{Gravwaves,Cavities} that we recapitulate below.

\begin{enumerate}
\item There should be some information about the measured observable which is encoded in the probe  system after the interaction. 
\item The measurement should not affect the measured observable  after the measurement. 
\item The measurement should be repeatable: Identical repeated measurements of the system  should consistently provide the same outcome.                                                                                                                                                                                                                                                                                                                                                                                                                                                                                                                                                                                                                                                                                                                                                                                                                                                                                                                                                                                                                                                                                                                                                                                                                                                                                                                                                                                                                                                                                                                                                                                              
\end{enumerate}

In our previous work we proposed to obtain information about the average photon population through an interferometric phase. If we already know that the target state is a Fock state, we can distinguish Fock states from each other. The mode invisibility technique is indeed a QND measurement for the following reasons:
\begin{enumerate}
\item The interferometric phase encodes information about the number of photons and it is resolvable to a precision that tells apart few-photon states satisfying the QND criterion 1.
\item The probability that the measurement takes the system to a  different state is many orders of magnitude below the intensity of the signal ($\mathrm{P\sim 10^{-22}}$ for physical parameters), showing that the system does not get perturbed after the measurement.
\item Given this probability for physical parameters, for our measurement outcome to be significantly altered, the measurement has to be repeated of the order on more than $\mathrm{10^{15}}$ times.
\end{enumerate}
 In this work we generalize the mode-invisibility technique to measure more general states of light such as coherent states and squeezed coherent states. However, it is not guaranteed \textit{a priori} whether the QND criteria are satisfied in this more general scenario. Nevertheless, we will show that the mode invisibility technique,  originally intended for the measurement of Fock states,  still provides a QND way to acquire information about general states of light. 

 \section{Background}
 
Let us briefly introduce the usual description of the different states of light to which we will apply the mode-invisibility technique.

First, a coherent state  of light  $\mathrm{\ket{\alpha}, ~(\alpha=|\alpha|e^{i \phi} \in \mathbb{C})}$ is defined as the eigenstate of the annihilation operator $\mathrm{(\hat{a})}$ when acting from the left (i.e., $\mathrm{\hat{a} \ket{\alpha} = \alpha \ket{\alpha}}$). It models relatively well the photon population of the outcome of a stationary laser source. In the number basis, $\mathrm{\{\ket{n}\}}$, a coherent state $\mathrm{\ket{\alpha}}$ takes  the form
\begin{align}\label{rep1}
\ket{\alpha} = & e^{-\frac{|\alpha|^{2}}{2}} \sum_{n=0}^{\infty}  \frac{\alpha^{n}}{\sqrt{n!}}\ket{n},
\end{align}
which is obviously well normalized $\mathrm{\langle \alpha | \alpha \rangle = 1}$ . It is straightforward to check that the expectation value of the photon number in a coherent state  is $\mathrm{\langle \hat{n} \rangle = |\alpha|^{2}}$. Additionally, it can be shown that  the action of the displacement operator 
\begin{equation}\label{dis}
D(\alpha) = \operatorname{exp}\Big(\alpha \hat{a}^{\dagger} - \alpha^{*}\hat{a}\Big),
\end{equation}
on the vacuum state is
\[\ket{\alpha}=D(\alpha)\ket0.\]
The action of the displacement operator on the creation operator $(a^{\dagger})$ and annihilation operator $(a)$ which obey the commutation relation $\mathrm{[a,a^{\dagger}] = 1}$, is given by 
\begin{subequations}
\begin{align}
D(\alpha)^{\dagger}\hat{a}D(\alpha) &= a + \alpha, \label{tone}\\
 D(\alpha)^{\dagger}a^{\dagger}D(\alpha) &= a^{\dagger} + \alpha^{*}.\label{ttwo}
 \end{align}
\end{subequations}
 respectively.

The second state of light we wish to discuss is the squeezed state \cite{ScullyBook}. Squeezed states describe very well the outcome of many non-linear quantum optics light sources (e.g, parametric down-conversion, etc.) \cite{sqcoherent1} and they have been widely studied in the past years as they have been proven useful for optical communications \cite{opticalcom,squeezedlight} and for quantum metrology \cite{Rideout2012, Gravwaves} among other disciplines. They are states that minimize the uncertainty relationships with phase-sensitive fluctuations (where the uncertainty in
the $\mathrm{x}$ and $\mathrm{p}$ quadratures is not necessarily the same).  
We will study the properties of squeezed states \cite{sqcoherent};   in particular squeezed coherent states and squeezed vacuum states \cite{PRL}. We will use the mode-invisibility technique to show how one can extract information about the degree of squeezing, the amplitude of the displacement, and some features of the Wigner function of squeezed coherent states of light ( the relative phase between squeezing and displacement)  without significantly perturbing the state. 

A  representation of a squeezed coherent state can be obtained by applying the squeeze operator
\begin{align}\label{sq}
S(\zeta)=\operatorname{exp}\Big(\frac{1}{2} \zeta^{*}\hat{a}\hat{a}-  \frac{1}{2} \zeta \hat{a}^{\dagger}\hat{a}^{\dagger} \Big), \qquad \zeta = r e^{i\phi},
\end{align}
on the coherent states of light
\[ \ket{\zeta,\alpha}\equiv S(\zeta) \ket{\alpha}.\]
 
$\mathrm{S(\zeta)}$ is a unitary operator $\mathrm{[ \hat{S}(\zeta)\hat{S}^{\dagger}(\zeta)=\hat{S}^{\dagger}(\zeta)\hat{S}(\zeta) = 1]}$ with magnitude $\mathrm{|S(\zeta)|=r,~(0\leq r < \infty)}$ and phase $\mathrm{\phi,~(0 \leq \phi \leq 2 \pi)}$ respectively \cite{squeezedlight}. The annihilation and creation operators transform under squeezeing as 
\begin{subequations}
\begin{align}\label{transformation}
b_{\kappa} = S(\zeta)^{\dagger}_{\kappa}a_{\kappa}S(\zeta) =&a_{\kappa} \cosh(r) - a^{\dagger}_{\kappa}e^{i\phi}\sinh(r),\\
b^{\dagger}_{\kappa}=S(\zeta)^{\dagger}_{\kappa}a^{\dagger}_{\kappa}S(\zeta)_{\kappa} =&a^{\dagger}_{\kappa} \cosh(r) - a_{\kappa}e^{-i\phi}\sinh(r).
\end{align}
\end{subequations}
Note that this transformation is canonical ($\mathrm{[b,b^{\dagger}]=1}$).   A QND measurement on a squeezed coherent state implies that 
\begin{align}\label{QNDsq}
\ket{\zeta,\alpha(T)} \simeq \ket{\zeta,\alpha (0)}
\end{align}
where we consider an interaction time from $\mathrm{t=0}$ to $\mathrm{t=T}$. Equation (\ref{QNDsq}) indicates that the state is unchanged after the measurement process. Quantitatively this criterion will manifest itself insofar as the transition probability of the system into a different state during the duration of the measurement is approximately zero. 

The mode-invisibility technique  measures the photon content in a given state of light  \cite{marvy2013}. How this non demolition measurement technique can be sensitive to the relative phase of the displacement operator $\mathrm{D(\alpha)}$ and squeeze operator $\mathrm{S(\zeta)}$ that characterize a squeezed coherent state does not seem obvious. We will show in this paper that we can indeed detect this relative phase difference without difficulty. We will also show that the interferometric method \cite{marvy2013} can achieve a squeezing sensitivity within the range $\mathrm{r \in [0,2]}$ that is experimentally realizable \cite{PhysRevLett.104.251102}.

\section{Setting}
\begin{figure}[h!]
\includegraphics[width=.55\textwidth]{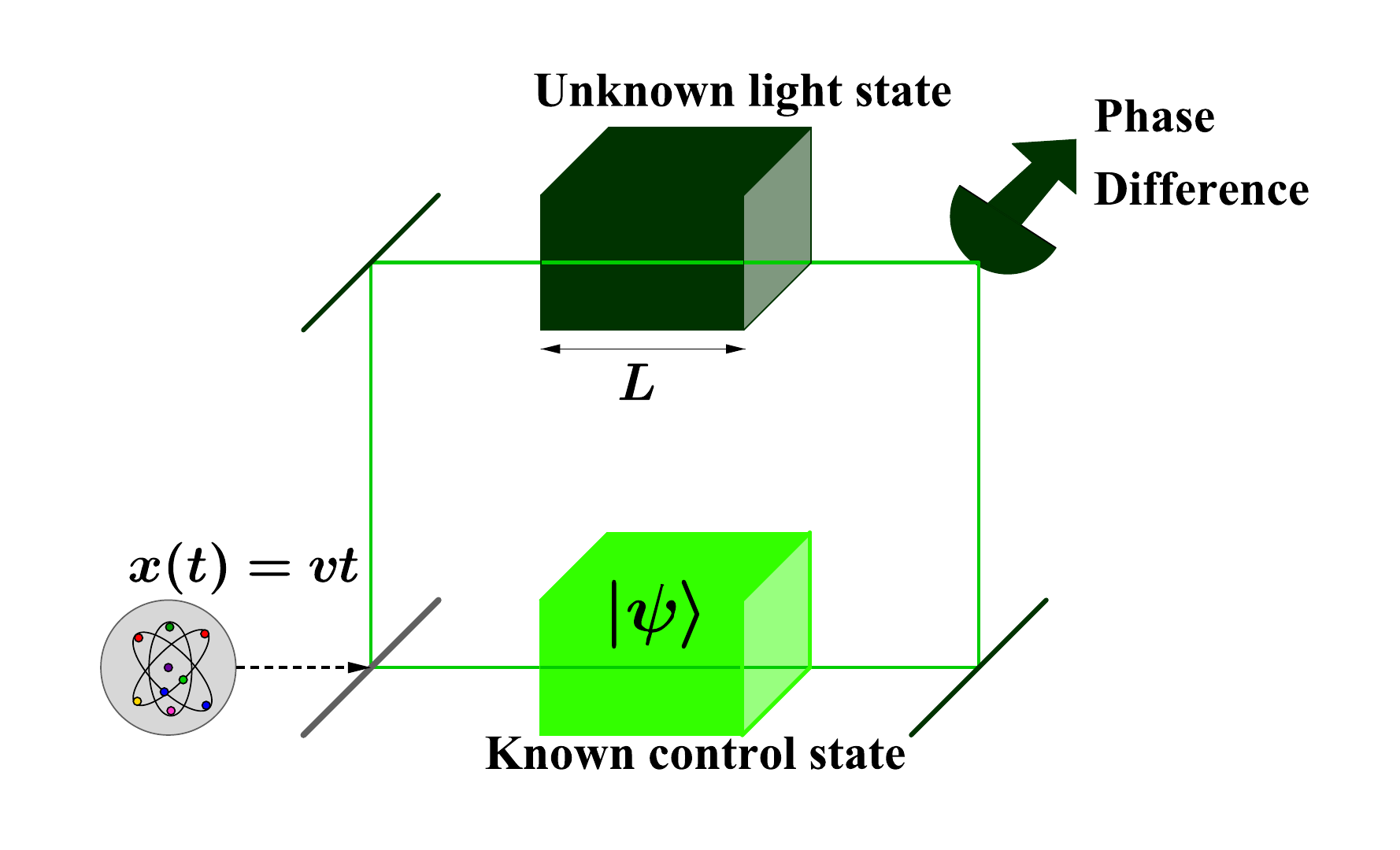}
\caption{(Color online)  Measurement setup to detect the properties of an unknown state of light trapped in a cavity using a known control light state in another cavity as a reference. A detector moving at constant $\mathrm{v}$ is sent into the interferometer to probe the unknown light state on resonance, but in a nondestructive manner using the mode- invisibility technique.}
\label{scheme}
\end{figure}
Our goal is to present a technique to measure the features that characterize coherent states and squeezed states (squeezed coherent states and squeezed vacuum states) of light without significantly perturbing them. We consider a  detector in the form of a single atom with two energy levels---ground $\mathrm{\ket{g}}$ and   excited $\mathrm{\ket{e}}$---with atomic transition frequency $\mathrm{\Omega}$. The detector is coupled to the target mode of the light field in an optical cavity. We will focus on the dynamical phase acquired by the atom crossing the cavity.  If its coupling to the field state is very weak, we  expect  that after an interaction time $\mathrm{T}$, the atom will exit the cavity without altering the probability distribution of the field state \cite{serge}, perhaps with a dynamical phase. We seek to  enhance the phase acquired by the atom as it flies through the cavity so as to improve prospects for the measurement of this
phase via interferometry.  If the atom is highly off resonant with the probed field mode, the action of that atom on the field will be minimized. However, the phase is also larger the closer we are to resonance  \cite{marvy2013}.  Ideally we would like to let the atom interact strongly with the relevant field mode, placing the field mode in resonance with the atomic gap. 

Figure \ref{scheme} shows our measurement setup  as illustrated in \cite{Unruh, Quantumthermometer, marvy2013}. In the present work we  analyze this basic setting when the target field state is an optical coherent  state or a squeezed state defined by the parameters $\mathrm{\alpha}$ and $\mathrm{\zeta}$ respectively. We show that the phase acquired by an atom crossing a cavity containing either of these quantum field states will carry information about their parameters without perturbing the field states significantly. By so doing we will be able to differentiate a coherent state with parameter $\mathrm{\alpha}$ from another with parameter $\mathrm{\alpha'}$. In the same way, we will also be able to  differentiate between a squeezed state with parameter $\mathrm{\zeta}$ and another with parameter $\mathrm{\zeta'}$. We will see that it is also possible to gain some information about the phase space distribution of a squeezed coherent state other than the expectation value of the photon number. Finally,  we will see that the mode invisibility technique using   simple coherent states  as an interferometric reference can be employed to probe different states of light (either Fock states, other coherent states, or squeezed states).  This class of states is generally easy to produce in the laboratory. 

For the light-atom interaction we consider the same  model as in  \cite{marvy2013}, regarding the atom as a two-level system, much like a Jaynes-Cummings Hamiltonian  in which neither the single-mode nor the rotating-wave approximation is carried out.  Specifically the light-matter interaction Hamiltonian \cite{Unruh, Quantumthermometer,AasenPRL} will be the Unruh--de Witt model $H_{I} = \lambda  \mu(t) \phi[x(t)]$, which is known   to model the interaction between two level atoms and the electromagnetic (EM) radiation field if there is no exchange of orbital angular momentum involved in the atomic transition \cite{Wavepackets,Alvaro}. The atom couples pointlikewise to the field  $\mathrm{\phi[x(t)]}$ along its trajectory $\mathrm{x(t)}$ through its monopole moment $\mathrm{\mu(t) = (\hat{\sigma}_{+} + \hat{\sigma}_{-})}$ where $\mathrm{\sigma^{+}}$ and $\mathrm{\sigma^{-}}$ are the creation and destruction operators for a two level atom respectively. $\lambda$ is the coupling strength. The quantum field is contained in a perfectly reflective cavity of length $L$. Under these assumptions, the interaction Hamiltonian is

 \begin{align}\label{interactionhamiltonian}\nonumber
H_{I} &=    \lambda\bigg( \sigma^{+}e^{i\Omega t} + \sigma^{-}e^{-i\Omega t}\bigg) \\
&\times \sum_{\kappa}\frac{1}{\sqrt{k_{\kappa}L}}\bigg(a_{\kappa}^{\dagger}e^{i\omega_{\kappa}t(t)} + a_{\kappa}e^{-i\omega_{\kappa}t(t)}\bigg)\sin(k_{\kappa}x),
\end{align}
where the field is expanded in terms of the stationary wave modes of the Dirichlet cavity. We define the time evolution operator from a time $t=0$ to a time $t=T$ as
\begin{align}\label{unitary}
U(0,T) = \mathcal{T}\text{exp}\bigg[-\ii\int_{0}^{T} dt H_{I}(t)\bigg],
\end{align}
where $\mathcal{T}$ is a time ordering. Under this evolution operator, the $n-$th order perturbative correction to the joint state of the atom-field system  will be given by 
\begin{align}\label{evolution}
\ket{\psi(T)^{n}} = U^{(n)}(0,T)\ket{\psi(0)},
\end{align}
after time $\mathrm{T}$, where $\ket{\psi(0)}$ is the initial joint state, with the different $U^{(n)}(0,T)$ being
\begin{equation}\label{eq:pert}
U(0,T) \!=\! \openone\!\underbrace{-\ii\!\!\int_{0}^{T}\!\!\!\!\!d t_1 H_{I}(t_1)}_{U^{(1)}}\underbrace{ -\!\!\int_{0}^{T}\!\!\!\!\!dt_1 \!\!\int_{0}^{t_1}\!\!\!\!\!dt_2\,    H_{I}(t_1) H_{I}(t_{2})}_{U^{(2)}}+\hdots
\end{equation}

\section{The measurement scheme}

\subsection{The weak adiabatic assumption and mode invisibility}\label{weaker}

Suppose the atom   is  prepared in its ground state $\mathrm{\ket{g}}$. To ensure that   its interaction leaves the field unperturbed, we will compute its transition probability into a state different from the original configuration, and require this quantity to be very small.  This is equivalent to demanding that the probability that the system remains in the same initial state after the atom crosses the cavity is approximately unity. In other words, we require
\begin{align}\label{hypo}
\big | \langle \psi (0) | U(0,T)| \psi(0) \rangle \big | ^{2} \approx 1;
\end{align}
hence the joint system evolves to the same system apart from a phase factor  
\begin{align}\label{phased}
|\psi(T) \rangle = U(0,T)|\psi(0) \rangle \approx e^{i \gamma}|\psi(0) \rangle,
\end{align}
where $\mathrm{\gamma}$ is the phase factor to be determined. We call this condition the ``weak adiabatic assumption''. As we will discuss below, this criterion will be satisfied for all the cases considered here due to mode invisibility. The weak adiabatic conditions \eqref{hypo} and \eqref{phased} ensure that the QND criteria 2 and 3 are satisfied.  

Since global phases are not measurable we will employ an atomic interferometric scheme \cite{marvy2013} to  compare the phase difference between two quantum field states:  the unknown quantum field state that we would like to measure, and a known quantum field state that acts as a reference.  This will allow us to  identify states within a known one-parametric family of states.
 
Consider the coupling between our detector and a squeezed coherent state trapped in an optical cavity. Let $\mathrm{\ket{\zeta, \alpha}_{\beta}}$ be a squeezed coherent state in the cavity mode $\mathrm{\beta}$ of frequency $\mathrm{\omega_\beta}$ while all the rest of the modes are prepared in very sparsely populated states. Therefore the joint initial state of our system will be well described as
\begin{align}\label{initialstate}
\ket{\psi(0)}=\ket{g}\otimes \ket{\zeta,\alpha}_{\beta}\bigotimes_{\beta \neq \gamma}\ket{0_\gamma}.
\end{align}

Using the evolution operator \eqref{unitary}, one can evaluate the state $\mathrm{\ket{\psi(T)}}$ of the system at a time $\mathrm{T}$, [see Eq. \eqref{evolution}]. Since we require that the interaction between the atom and the squeezed coherent light does not alter the state of the combined system, we can estimate the excitation probability of the combined system after an interaction time $T$ to be  \cite{marvy2013},
\begin{align}\label{probb}
P_{\ket{e}} = \langle e|  \operatorname{Tr}_{F}\Big[U^{(1)} \rho(0) U^{(1)\dagger} \Big] \ket{e}
\end{align} 
where $\mathrm{U^{(1)}}$ is the first order contribution to the unitary operator \eqref{unitary}, given as
\begin{align}\label{first}
 U^{(1)} = -\ii\sum_{\kappa} \frac{1}{\sqrt{k_{\kappa}L}}\Big( \sigma^{+}a_{\kappa}^{\dagger}I_{+,\kappa} +  
+ \sigma^{+}a_{\kappa}I_{-,\kappa}^{*}\Big)
\end{align}
and for notational convenience we have defined
\begin{align}\label{prob}
I_{\pm, \kappa} = \int_{0}^{T} \mathrm{d}te^{\ii(\pm \Omega + \omega_{\kappa})t}\sin[k_{\kappa}x(t)].
\end{align}
$\mathrm{\operatorname{Tr}_{F}}$ is the partial trace of the joint system over the squeezed-coherent field state. In the form of a density operator, Eq. \eqref{initialstate} gives $\mathrm{\rho_{0} = |g\rangle \langle g| \otimes | \zeta,\alpha\rangle_{\beta} \langle \zeta, \alpha|_{\beta}}$. If we define this density operator in terms of the squeeze and displacement operators it is easy to evaluate Eq. \eqref{probb} following the cyclic property of a trace (see the Appendix). After an evolution time $\mathrm{T}$ we obtain
\begin{align}\label{tras} \nonumber
P^{\alpha,r}_{\ket{e}} =\lambda^{2} \Bigg[\frac{1}{k_{\beta}L} ( | I_{-,\beta}|^{2} + | I_{+,\beta}|^{2}) \big(C^{2}(r) + S^{2}(r)\big)|\alpha|^{2}\\\nonumber
 - \frac{2}{k_{\beta}L} (| I_{-,\beta}|^{2} + | I_{+,\beta}|^{2}) S(r) C(r) \operatorname{Re}\big[| \alpha|^{2} e^{ \ii(2\theta - \phi)} \big]\\
+
( |I_{-,\beta}|^{2} + |I_{+,\beta}|^{2})\frac{S^{2}(r)}{k_{\beta}L}  + \sum_{\gamma}\frac{|I_{+,\gamma}|^{2}}{k_{\gamma}L}\Bigg],
\end{align}
where, to shorten notation, we have defined $\mathrm{\sinh(r) = S(r)}$ and $\mathrm{\cosh(r) = C(r)}$. 

 For a large range of values  of $\mathrm{r, \theta, \phi, \alpha}$ (and in particular for all those studied in detail below),  and for the speed of the atomic probe in the range $\mathrm{v\approx 1-1000 m/s}$, we find that we can achieve  values  $\mathrm{P^{\alpha,r}_{\ket{e}} \ll 1}$,   implying that the  detector negligibly perturbs the field state during the interaction as per   QND requirement 2. Using \eqref{dis} and  \eqref{sq}  we see that   the squeezed coherent state $\mathrm{\ket{\zeta, \alpha}_{\beta} = S(\zeta)_{\beta}D(\alpha)_{\beta}\ket{0}}$ contains both  the squeezed vacuum state $\mathrm{S(\zeta)\ket{0}}$ and  the coherent state $\mathrm{D(\alpha)\ket{0}}$ as special cases.  Hence  \eqref{tras} also gives   the correct transition probabilities for the squeezed vacuum and the coherent state.  Namely, for a coherent state,

\begin{align}\label{cohtrans}
P^\alpha_{\ket{e}} =\Bigg[\frac{|\alpha|^{2}}{k_{\beta}L} ( | I_{-,\beta}|^{2} + | I_{+,\beta}|^{2}) +
\sum_{\gamma}\frac{|I_{+,\gamma}|^{2}}{k_{\gamma}L}\Bigg],
\end{align}
and for the squeezed vacuum,
\begin{align}\label{sqtras} 
P^r_{\ket{e}} =\Bigg[\frac{S^{2}(r)}{k_{\beta}L}  
( |I_{-,\beta}|^{2} + |I_{+,\beta}|^{2})+ \sum_{\gamma}\frac{|I_{+,\gamma}|^{2}}{k_{\gamma}L}\Bigg].
\end{align}

As shown in equations \eqref{tras}, \eqref{cohtrans} and \eqref{sqtras} respectively, the transition probabilities $\mathrm{\ket{P}_{\ket{e}}}$ in all cases depend on the intergals $\mathrm{|I_{\pm,\beta}|^{2}}$ with $\mathrm{I_{\pm,\beta}}$ defined as in Eq. \eqref{prob}. This integral (the rotating-wave resonant term) gives by far, the largest contribution to the probability of transition of the system. Recall that when the atomic probe crosses the cavity at constant (and non-relativistic) speed $\mathrm{x(t)=vt}$ the integral \eqref{prob} is easily evaluated to give  \cite{marvy2013}, 
\begin{align*}
I_{\pm, \beta} =   \frac{\left[e^{\ii\frac{L}{v}(\omega_\beta\pm\Omega )}(-1)^{\beta} -1\right] L v\sqrt{\beta\, \pi} }{\left(\beta\, \pi v\right)^2-L^2(\omega_\beta\pm\Omega)^2}.
\end{align*}
At resonance $\mathrm{\omega_{\beta} = \Omega}$,  we have that the leading-order contribution to the transition amplitude is
\begin{align*}
I_{-, \beta} =   \frac{\left[(-1)^{\beta} -1\right] L }{(\beta\, \pi)^{3/2} v}.
\end{align*}
 Previously  we demonstrated that this term could be canceled \cite{marvy2013}. From the interaction Hamiltonian \eqref{interactionhamiltonian}, we see that multiples of the second harmonic  ($\mathrm{\beta=2,4,6,\dots}$) have the property that their spatial wave functions are of odd parity while for the  rest of the harmonics they are of even parity.  Thus an atom that flies through the cavity will be effectively changing the sign of the coupling (in the first order correction terms)   an even number of times.  Hence although the detector may be instantaneously altering the field state along its trajectory,  if it travels at constant speed  and probes only the even cavity modes, the overall effect on the field state when considering the whole trajectory will be negligible.

 Upon eliminating this largest term,   the maximum contribution to the transition probability is from the sum of all the $\mathrm{|I_{+,\beta}|^{2}}$ terms, which is convergent.   For the set of parameters studied in this paper, we find that the transition probability for an atom crossing a cavity sustaining a coherent state is of the order of $\mathrm{\sim10^{-22}}$ multiplied by $\mathrm{|\alpha|^{2}}$ [which is easy to evaluate from \eqref{cohtrans} noting that the integrals $\mathrm{I_{\pm}}$ were already evaluated in \cite{marvy2013}]. This indicates that our measurement technique is indeed a QND measurement scheme, allowing us to keep the `weak adiabatic assumption' for relatively large values of the parameters $\mathrm{|\alpha|}$ and $\mathrm{r}$.  
 

Once we have made sure that the weak adiabatic condition is satisfied, in \eqref{tras}, \eqref{cohtrans}, and  \eqref{sqtras}, we can proceed to calculate the phase acquired by the atom upon interaction with each light field. To this end we need to compute the leading-order contribution to the phase.

\subsection{Calculating the second order contribution to time evolution}
The second order contribution to the time evolution of the atom-field system  is given by the expression
\begin{align}\label{fr}
\ket{\psi^{(2)}}= U^{(2)}\ket{\psi(0)},
\end{align}
where $\mathrm{U^{(2)}}$ is the second order contribution to the unitary operator \eqref{unitary}. 
\begin{align}\label{evol2}
&U^{(2)}=-\sum_{\kappa, \delta}\int_{0}^{\frac{L}{v}}dt\int_{0}^{t}dt'H_{I}(t)H_{I}(t')
\end{align}
We prepare the atom initially in its ground state. Hence, the $(\sigma^{+}\sigma^{-})$ terms in $U^{(2)}$  give  zero contribution to Eq. \eqref{fr}. The only surviving terms are proportional to  $(\sigma^{-}\sigma^{+})$. After a lengthy calculation (see the Appendix ), we obtain the leading-order correction to the phase factor for the squeezed coherent state to be,
\begin{align}\label{order}\nonumber
&\ket{\psi^{(2)} }=- \lambda^{2} \Bigg( \sum_{\gamma}\frac{C_{+,\gamma}^{*}}{k_{\gamma}L} + (C_{+,\kappa}^{*}+ C_{-,\kappa})\frac{S^{2}(r)}{k_{\kappa}L} \\\nonumber
& - \frac{2}{k_{\kappa}L} (C_{+,\kappa}^{*}  + C_{-,\kappa})S(r) C(r) \operatorname{Re}\big[| \alpha|^{2} e^{ \ii(2\theta - \phi)} \big] \\ &+\frac{|\alpha|^{2}}{k_{\kappa}L}  [C^{2}(r) + S^{2}(r)] (C_{+,\kappa}^{*} + C_{-,\gamma})\Bigg) \ket{\psi(0)} + \ket{\psi(T)}_{\perp},
\end{align}
 where
\begin{align}\label{note}
C_{\pm,\kappa} &= \int_{0}^{L/v}dt \int_{0}^{t}dt' e^{\ii(\omega_{\kappa} \pm \Omega)(t-t')} \sin(k_{\kappa}vt)\sin(k_{\kappa}vt');
\end{align}
 $\mathrm{\ket{\psi(0)}}$ is the initial state of the joint atom--squeezed coherent system, and $\mathrm{\ket{\psi(T)}_\perp}$  is the second-order correction which is orthogonal to the initial state and which is irrelevant to the computation of the phase.  $\mathrm{\ket{\psi(T)}_\perp}$, should be small enough for all our assumptions to hold. Its magnitude will have an impact on the visibility of the fringes in the interferometric experiment as detailed in \cite{marvy2013}. Since this constitutes a sub leading  contribution to the probability of modifying the state $\mathcal{O}(\lambda^2)$ in amplitudes, $\mathcal{O}(\lambda^4)$ in the probabilities) that is not canceled by the mode invisibility technique, as opposed to the leading order, its magnitude must also be kept under control  to guarantee the QND nature of the measurement. We will see that this term remains under control for all the relevant cases studied here.

 For $\mathrm{r= 0}$ we recover the case of a coherent state in the cavity mode we want to probe. For that particular case the expression simplifies to
\begin{align*}
\ket{\psi(T)^{(2)} }= -\lambda^{2}\Bigg(\frac{( C_{+,\kappa}^{*}+ C_{-,\kappa})}{k_{\kappa}L}|\alpha|^{2}  +
  \sum_{\gamma}\frac{C_{+,\gamma}^{*}}{k_{\gamma}L}\Bigg) \ket{\psi(0)} \\ + \ket{\psi(T)}_{\perp}
\end{align*}
Similarly for the case $\mathrm{\alpha = 0}$ we recover the squeezed vacuum. In this case the leading-order contribution is
\begin{align*}
\ket{\psi(T)^{(2)}}=-\lambda^{2}\bigg(\frac{( C_{+,\kappa}^{*} + C_{-,\kappa})}{k_{\kappa}L} S^{2}(r) + \sum_{\gamma}\frac{C_{+,\gamma}^{*}}{k_{\gamma}L}\bigg) \ket{\psi(0)} \\
+ \ket{\psi(T)}_{\perp}
\end{align*}

\subsection{Calculating the phase factor acquired by an atom after an interaction with a given light field}
According to our weak adiabatic scheme explained in Sec. \ref{weaker}, we expect that a state that starts at time $\mathrm{t=0}$ will evolve to almost the same state but for an additional global phase factor $\mathrm{\gamma}$ [see Eq. \eqref{phased}]. Upon simplification, Eq. \eqref{phased} reduces to
\begin{align}\label{phaze}
\eta = - \ii \operatorname{ln}\Bigg[ 1 - \lambda^{2}\langle \psi(0)| U^{(2)}(0,T)| \psi(T) \rangle \Bigg],
\end{align}
where $\mathrm{\eta}$ could be a complex number due to the presence of orthogonal terms in the leading order contribution to $\mathrm{\gamma}$, implying that $\mathrm{\gamma = \operatorname{Re}[\eta]}$. Multiplying Eq. \eqref{order} from the left by $\langle \psi(0)|$ and substituting back into Eq. \eqref{phaze} we find
\begin{align}\label{squeezecoherentphase}\nonumber
\eta(r,\alpha)& = -\ii \operatorname{ln}\Bigg[1 - \lambda^{2}\bigg( \frac{ \big(C_{+,\kappa}^{*}+C_{-,\kappa}\big)}{k_{\kappa}L} S^{2}(r)  + \sum_{\gamma}\frac{C_{+,\gamma}^{*}}{k_{\gamma}L}\\\nonumber
&+\frac{C_{+,\kappa}^{*} + C_{-,\kappa} }{k_{\kappa}L}  [C^{2}(r) + S^{2}(r)]|\alpha|^{2}\\
& -\frac{ 2}{k_{\kappa}L} (C_{+,\kappa}^{*}+ C_{-,\kappa}) S(r) C(r) \operatorname{Re}\big[| \alpha|^{2} e^{ \ii(2\theta - \phi)} \big] \bigg)\Bigg]
\end{align}
for the phase acquired by an atom flying through a cavity with a squeezed coherent state of light.

For the particular case  when $\mathrm{r= 0}$, we have the phase acquired by an atom crossing a cavity with coherent state sustained in it to be
\begin{align}\label{roguephase}
\eta(\alpha) &= -\ii \operatorname{ln}\Bigg[1 - \lambda^{2}\Bigg(\sum_{\gamma}\frac{C_{+,\gamma}^{*}}{k_{\gamma}L}+\frac{ (C_{+,\kappa}^{*}+C_{-,\kappa})}{k_{\kappa}L}   |\alpha|^{2}\Bigg)\Bigg].
\end{align}

 Similarly, when $\mathrm{\alpha = 0}$, the phase acquired by an atom crossing a cavity containing a squeezed vacuum state is
\begin{align}
&\eta(r) = -\ii \operatorname{ln}\Bigg[1 - \lambda^{2}\Bigg(\sum_{\gamma}\frac{C_{+,\gamma}^{*}}{k_{\gamma}L}+\frac{S^{2}(r) }{k_{\kappa}L}(C_{+,\kappa}^{*}+C_{-,\kappa})\Bigg)\Bigg].
\end{align}

\begin{figure*}[t]\begin{tabular}{ccc}
\includegraphics[width=.32\textwidth]{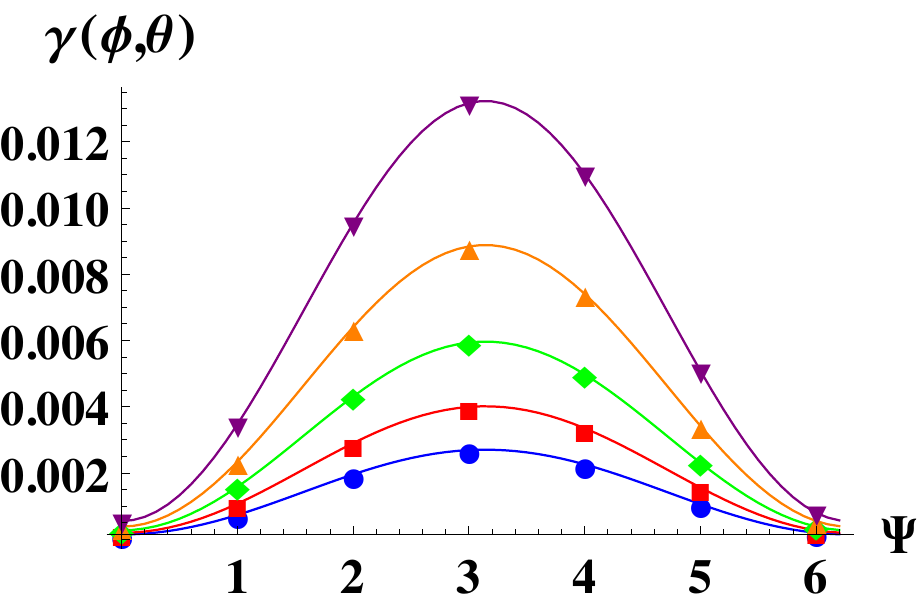}&
\includegraphics[width=.32\textwidth]{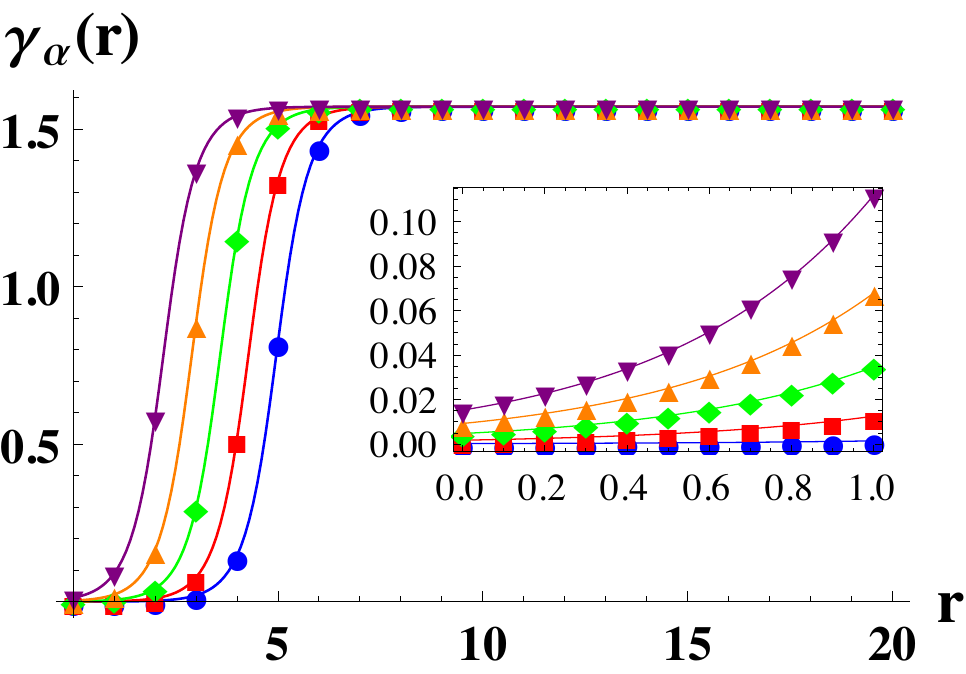}&
\includegraphics[width=.32\textwidth]{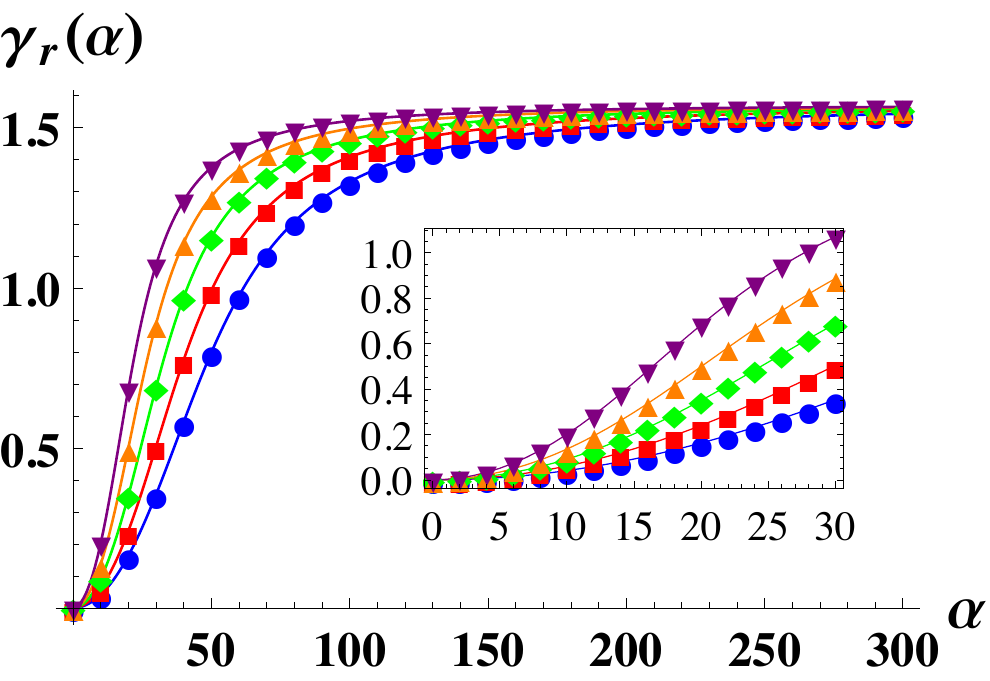}\end{tabular}
\caption{(Color online) Phase acquired by an atom crossing a cavity with squeezed coherent field trapped inside, from left to right: Left: as a function of the relative phase difference $\mathrm{\Psi}$ between the coherent operator and the squeezed operator; middle: as a function of the amplitude $\mathrm{r}$ of the squeeze operator $\mathrm{S(r, \phi)}$; right: as a function of the amplitude  $\mathrm{|\alpha|}$ of the coherent state operator $\mathrm{D(\alpha)}$. Each figure show the sensitivity of the phase to the state parameters. The physically realizable squeeze parameter range is shown in the inset in the middle figure. We see that in that regime reasonable sensitivity to the phase is attained.  However for large squeeze amplitudes $\mathrm{r>5}$, the phase value ceases to be sensitive to the squeeze amplitude.}
\label{phase-mag}
\end{figure*}

 Figure \ref{phase-mag} shows different plots of the measurable phase $\mathrm{\gamma}$ 
[the real value of $\mathrm{\eta(r,\alpha)}$] as a function of its dependent variables $\mathrm{\alpha} and \mathrm{r}$ and  the relative phase difference $\mathrm{\Psi = 2\theta - \phi}$. The atom is taken to fly through the cavity for a  time $\mathrm{t = {\beta \pi}/{\omega_{\beta}}}$ at a velocity $\mathrm{v=1000 m/s}$.  The inset in the center plot of Fig. \ref{phase-mag} shows that our method is sensitive to values of $\mathrm{r}$ that are experimentally realizable \cite{PhysRevLett.104.251102}, although we have plotted our results for values of $\mathrm{r}$ well beyond this range. We clearly see that while the probability that the system evolves to some different state is rendered negligible, the global phase factor (which can be determined via atomic interferometer by comparison with a known state) is not negligible.

\begin{figure}[h!]
\includegraphics[width=.40\textwidth]{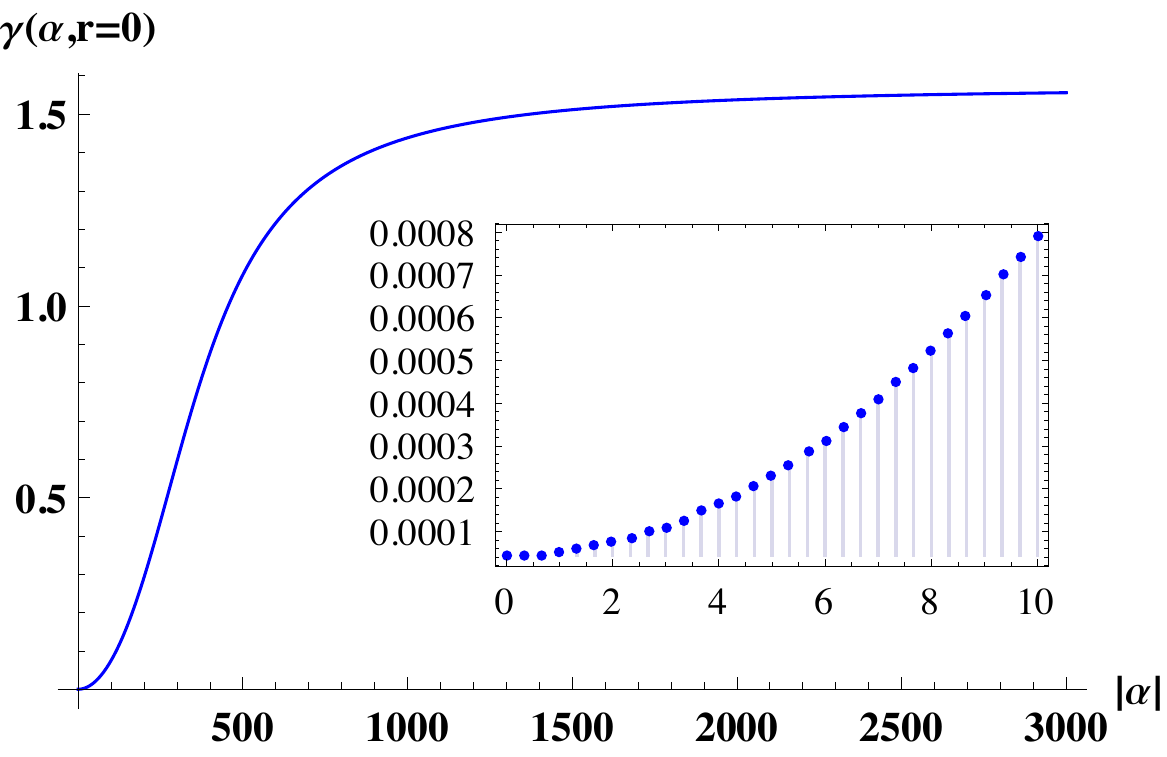}
\caption{(Color online) { Phase acquired by an atom that interacts with a coherent state as a function of the coherent state amplitude $\mathrm{|\alpha|}$. The inset shows an exponential curve for small values of $\mathrm{|\alpha|}$. For large values, the curve plateaus, making it increasingly difficult to gain information about $\mathrm{|\alpha|}$.}}
\label{comparecoherent1}
\end{figure}

\begin{figure}[h!]
\begin{center}
\includegraphics[width=.40\textwidth]{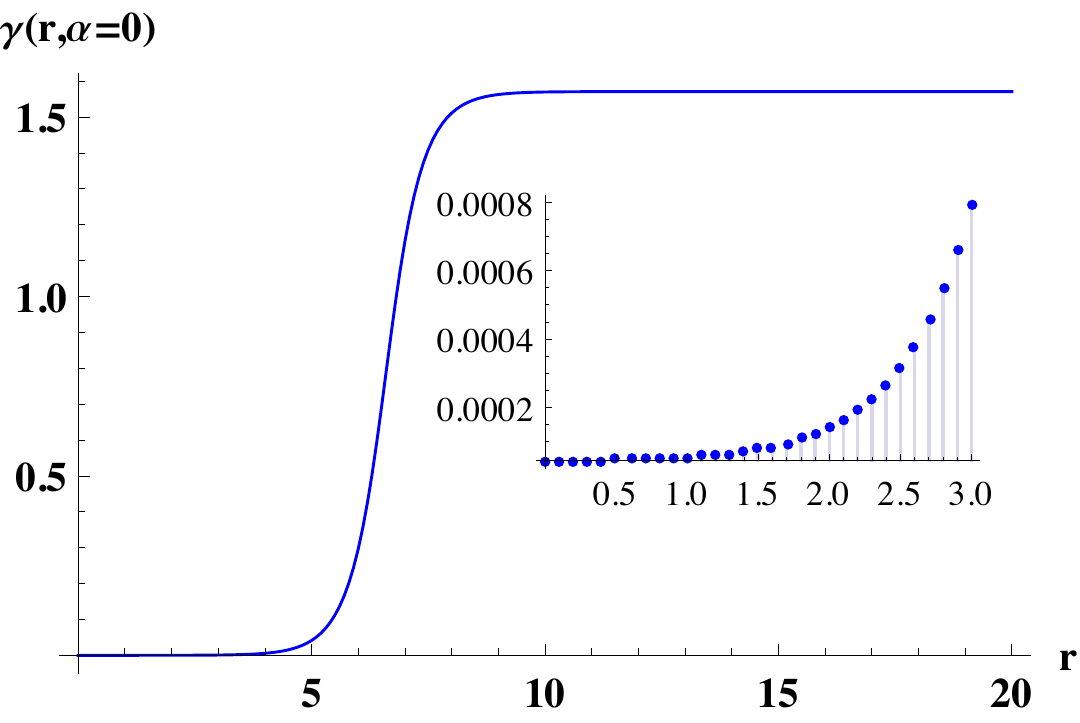}
\caption{(Color online) { Phase acquired by an atom that interacts with a coherent state as a function of the amplitude of the squeeze operator $\mathrm{r}$. The inset shows an exponential curve for small values of $\mathrm{r}$.  The curve eventually plateaus, making it increasingly difficult to gain information about $\mathrm{r}$. }}
\label{comparesqueezing}
\end{center}
\end{figure}

Figure \ref{comparecoherent1} shows a plot of $\mathrm{\operatorname{Re}[\eta(\alpha)]}$ as a function of the amplitude of the coherent state $\mathrm{|\alpha|}$ and Fig. \ref{comparesqueezing}  as a function of $\mathrm{r}$. 

\section{Phase sensitivity to the state parameters}

\begin{figure*}[t]\begin{tabular}{ccc}
\includegraphics[width=.32\textwidth]{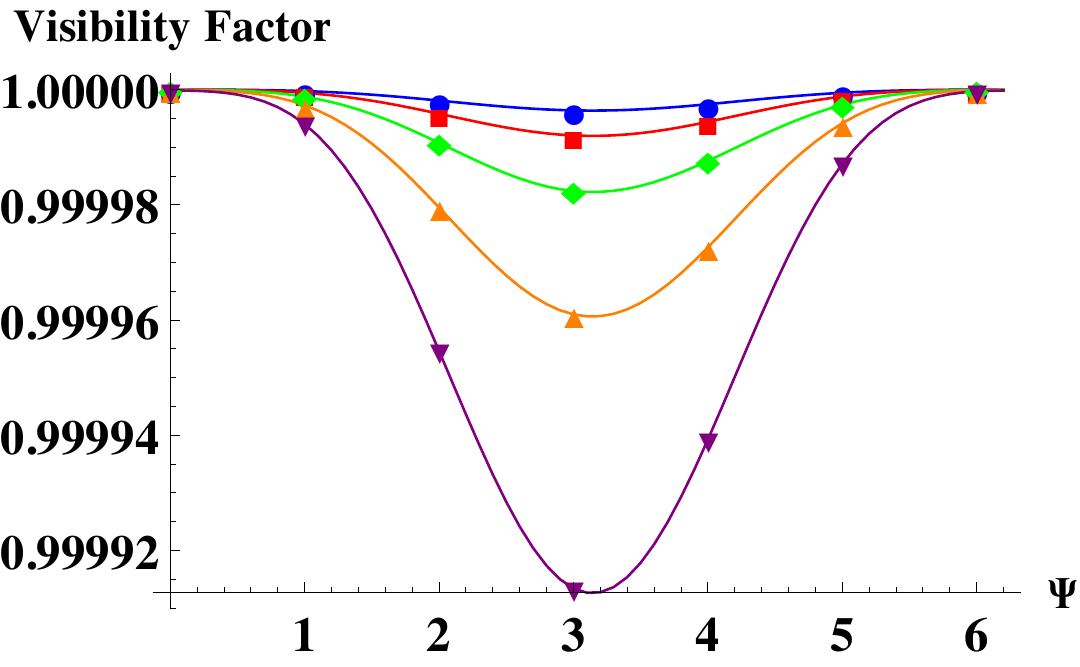} \includegraphics[width=.32\textwidth]{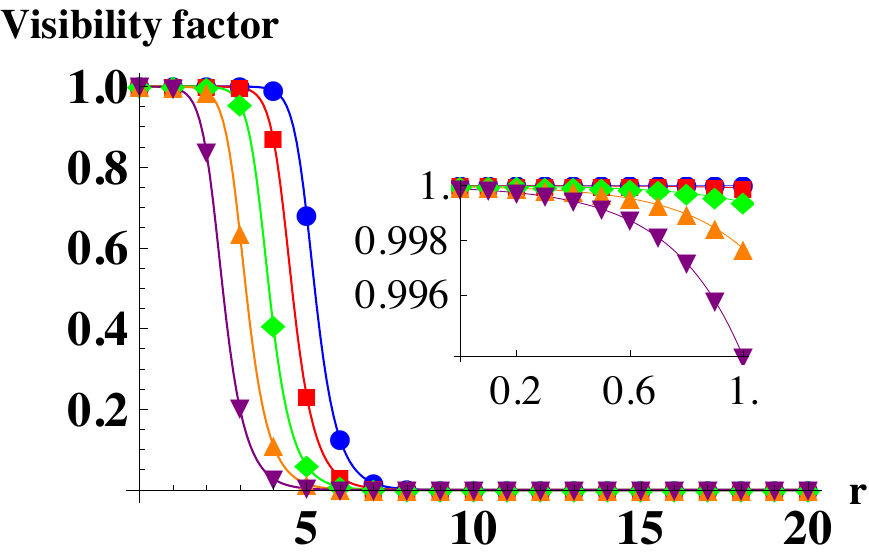}\includegraphics[width=.32\textwidth]{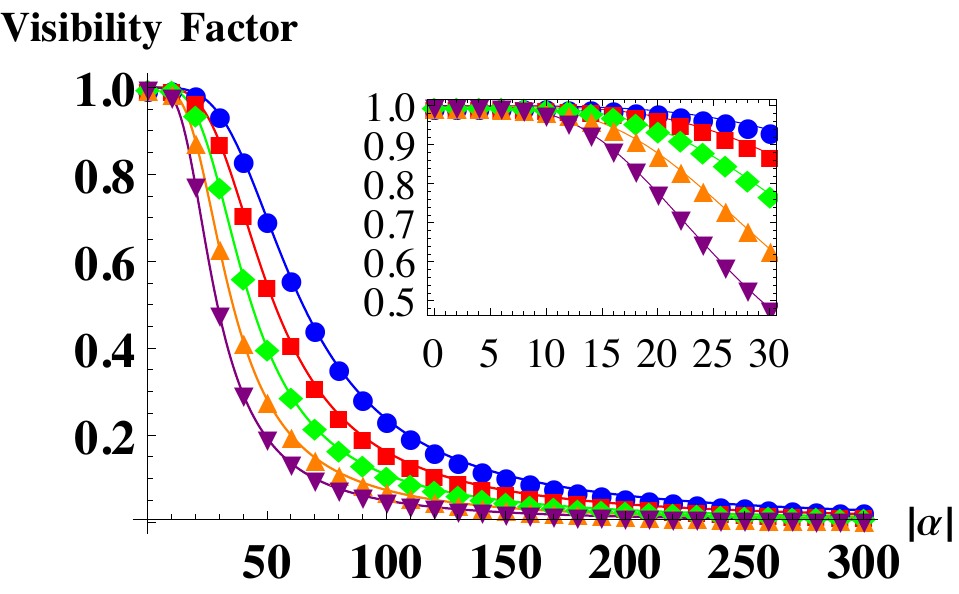}\end{tabular} 
\caption{(Color online): Visibility factor for the cases plotted in Fig. \ref{phase-mag}, from left to right. Left: as a function of the relative phase difference $\mathrm{\Psi}$ between the coherent operator and squeezed operator; middle: as a function of amplitude $\mathrm{r}$ of the squeeze operator $S(r, \phi)$; right: as a function of the amplitude  $\mathrm{|\alpha|}$ of the coherent state operator $D(\alpha)$. } 
\label{visibility_squeezestate}
\end{figure*}

Our final goal   is to be able to compare the different dynamical phases acquired by different states through an atomic interferometry experiment. For that we   need to know how the global phase acquired depends on the parameters of the relevant states studied.

From \eqref{squeezecoherentphase}  we see that the global phase acquired by the atom crossing a cavity where a squeezed coherent state is prepared, is sensitive to (1) the  magnitude of the displacement parameter $\mathrm{|\alpha|}$, (2) the  squeezing parameter $\mathrm{r}$, and (3) the relative angle $\mathrm{2\theta-\phi}$ between the squeezing  and  displacement operations in phase space. This means that the global phase contains information about the Wigner function of the squeezed coherent state: both the average number of photons and the relative phase between the squeeze and displacement operators. If we were able to measure it we could use it to characterize a coherent state or a squeezed state or to measure the direction of squeezing in phase space relative to the direction of displacement in a coherent squeezed state.

Figure \ref{phase-mag} shows how the dynamical phase is sensitive to these three parameters. For the relative phase between the squeezing and the displacement, we see that there is indeed an appreciable phase difference  that increases when the photon number expectation of the mode increases.
 
The figure also shows that the phase is extremely sensitive to the values of the squeezing parameter and the displacement parameter for a range of values (although it loses sensitivity when the expected photon population increases). This suggests that the measurement of this phase would be an extremely good method for probing, in a nondemolition way, states that are very scarcely populated, losing sensitivity as the photon population increases.

Figure \ref{visibility_squeezestate} shows the different visibility factors that would impact an interferometric experiment, where we compare the phase acquired by an atom going through a cavity with the target state of light and some other reference phase. Consistently with the previous results, we see that the method works much better in probing states with a low photon number expectation.

\section{Measurement settings}

We have determined how the global phase acquired by an atomic probe behaves as a function of the parameters of the probed states of light in a quantum non-demolition measurement setting. However global phases cannot be directly measured. We will therefore always need to use an interferometric scheme, such as the one described in Fig. \ref{scheme}, to obtain the phase acquired from the probed state relative to the phase of a known reference state.

When it comes to choosing a reference state, we would like it to be a state that is easy to prepare and control in the laboratory. In our previous work \cite{marvy2013} we used Fock states as reference states to probe unknown Fock states. However, this is less than ideal since the preparation and control of Fock states of light is a rather challenging enterprise in quantum optics \cite{ Hennrich2000}. We would like to remove this  constraint from the `mode-invisibility' QND measurement scheme. A natural candidate for a reference state is a coherent state. These states are among the easiest to prepare and control in quantum optic laboratories \cite{ScullyBook}. We have shown that the mode-invisibility technique is also applicable to coherent states of light.

Using a cavity with a coherent state sustained in it as a reference cavity (see the measurement setup \cite{marvy2013}) we shall see that it is possible to measure the expected photon content of Gaussian states and the relative phase between a squeeze and a displacement operation. Furthermore, we will see that it is possible to obtain the same measurement resolution as in \cite{marvy2013} by using a coherent state as reference instead of another controlled Fock state.

\subsection{Interferometric measurement of a squeezed coherent state using another coherent state as reference}

\begin{figure}[h!]
\begin{center}
\includegraphics[width=.40\textwidth]{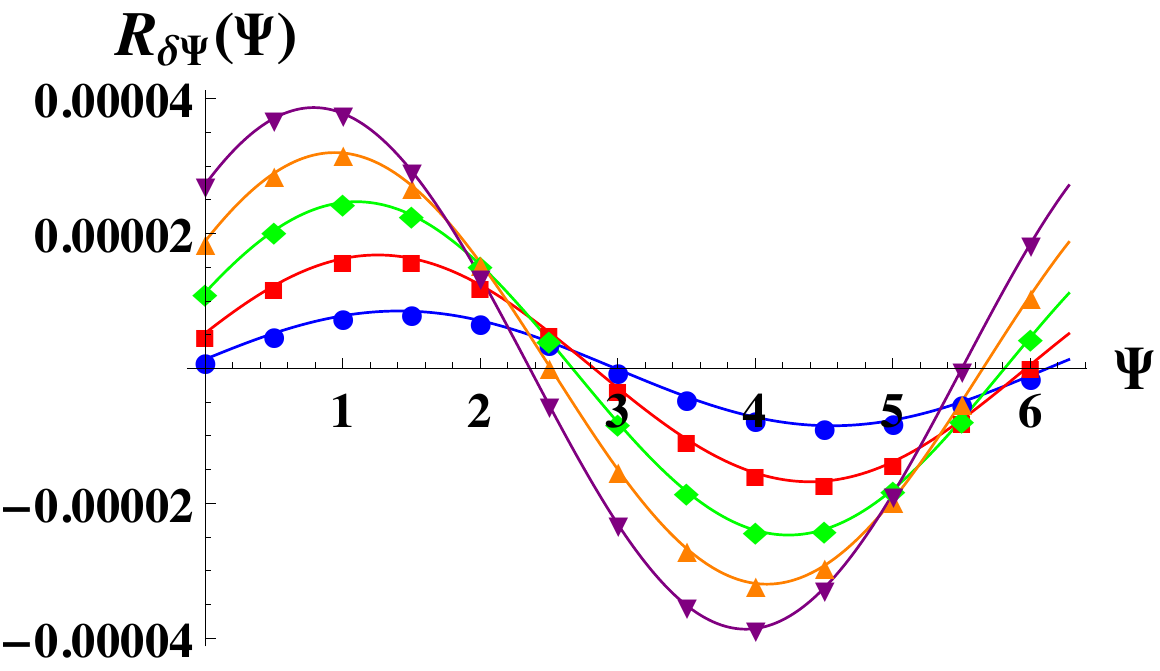}
\caption{(Color online): {  Phase resolution to distinguish the relative phase difference acquired by an atom that interacts with two different squeezed coherent states. Here we plot the relative phase difference between $\mathrm{\Psi}$ and $\mathrm{\Psi + \delta \Psi}$ for each squeezed coherent state respectively. Different curves with $\mathrm{\delta \Psi = 0.5 \pi, 0.4 \pi, 0.3 \pi, 0.2\pi}$ and $\mathrm{0.1\pi}$ are shown for the values of $\mathrm{|\alpha| = r = 1}$. }}
\label{comparism}
\end{center}
\end{figure}

The phase acquired by an atom crossing a cavity sustaining a squeezed coherent state of light is  given  in Eq. \eqref{squeezecoherentphase}. To measure this phase, we will compare it with that acquired by an atom crossing a cavity with a coherent state in it. This phase difference is given by the expression
\begin{align*}
\Delta\gamma(r,\alpha,\Psi) = \operatorname{Re}[-\ii \operatorname{ln} \chi(\alpha,r,\Psi)]
\end{align*}
where
\begin{align}\label{reff}\nonumber
\chi(\alpha,r,\Psi)&=\Bigg[1 - \bigg( \frac{ \big(C_{+,\kappa}^{*}+C_{-,\kappa}\big)}{k_{\kappa}L} S^{2}(r)  + \sum_{\gamma}\frac{C_{+,\gamma}^{*}}{k_{\gamma}L}\\\nonumber
&+\frac{C_{+,\kappa}^{*} + C_{-,\kappa} }{k_{\kappa}L}  (C^{2}(r) + S^{2}(r))|\alpha|^{2}\\\nonumber
& -\frac{ 2}{k_{\kappa}L} (C_{+,\kappa}^{*}+ C_{-,\kappa}) S(r) C(r) \operatorname{Re}\big[| \alpha|^{2} e^{ \ii(2\theta - \phi)} \big] \bigg)\Bigg] \\ &\times\Bigg[ 1\! -\!  \bigg(\frac{(C_{+,\kappa}^{*} + C_{-,\kappa})}{k_{\kappa}L}|\alpha_{R}|^{2}+\! \sum_{\beta}  \frac{C_{+,\beta}^{*}}{k_{\beta}L}\bigg)\Bigg]^{-1}
\end{align}
and we have defined $\mathrm{\alpha_{R}}$ as the amplitude of the reference coherent state. We show in Fig. \ref{comparism} a plot of the phase resolution to distinguish the relative phase difference acquired by an atom that interacts with a squeezed coherent state with $\mathrm{\Psi}$ from another with $\mathrm{\Psi + \delta \Psi}$.

\begin{figure}[h!]
\begin{center}
\includegraphics[width=.40\textwidth]{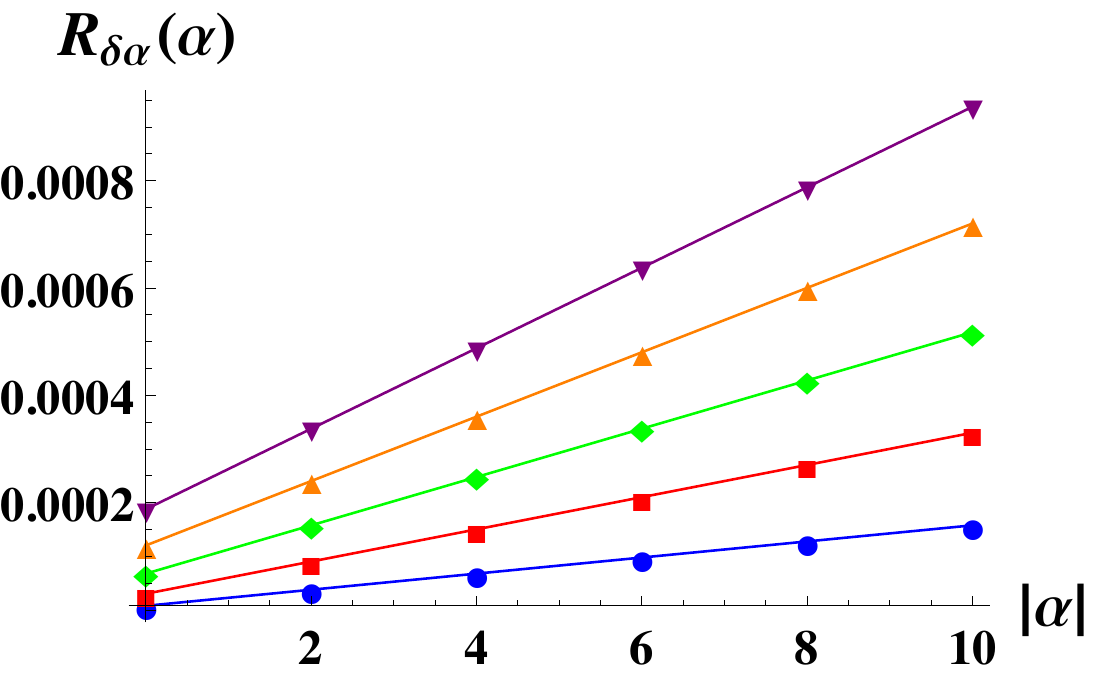}
\caption{(Color online): Phase resolution required to distinguish between a coherent state with amplitude $\mathrm{|\alpha|}$ and another coherent state with amplitude $|\alpha + \delta \alpha|$. We show the plots for values of $\mathrm{\delta \alpha = 1, 2, 3, 4}$ and  $\mathrm{5}$ respectively.}
\label{comparecoherent}
\end{center}
\end{figure}

\begin{figure}[h!]
\begin{center}
\includegraphics[width=.40\textwidth]{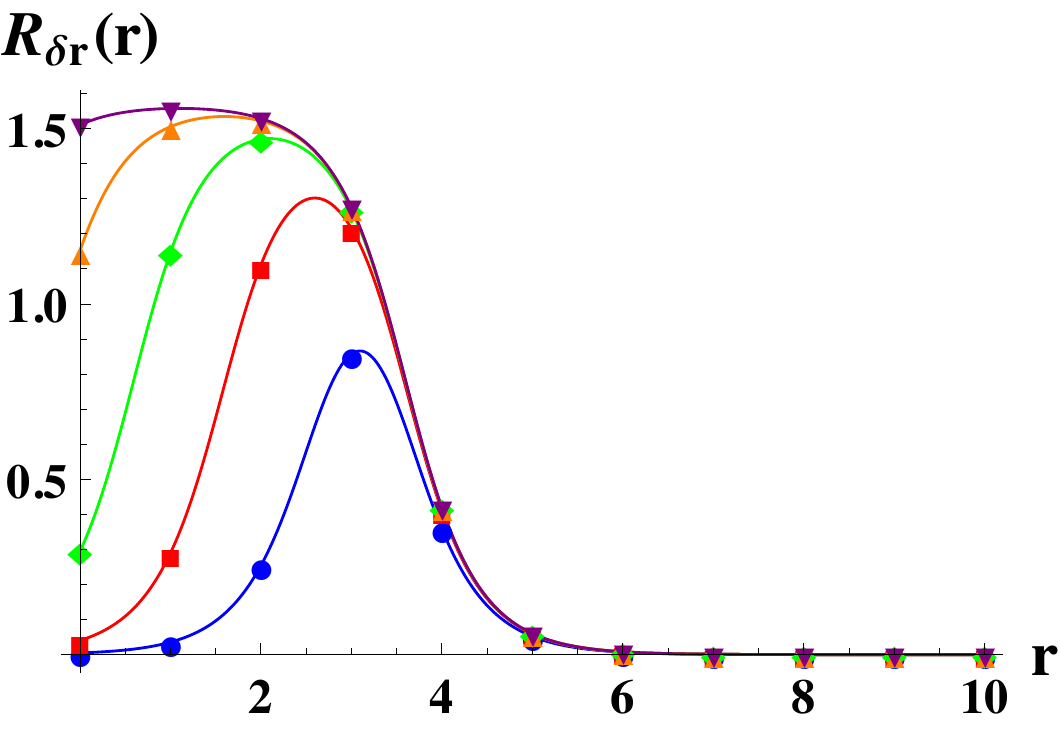}
\caption{(Color online) Phase resolution required to distinguish between a squeezed vacuum state with amplitude $\mathrm{r+\delta r}$ and another squeezed vacuum state with amplitude $r$. The difference curves are shown for $\mathrm{\delta r =1, 2, 3, 4}$ and  $\mathrm{\delta r = 5}$ respectively.  We see that we have more than enough resolution to differentiate between the amplitudes of two squeezed coherent states.}
\label{coherency}
\end{center}
\end{figure}

{ We can thus obtain the phase resolution required to distinguish between two coherent states and two squeezed vacuum states by setting $\mathrm{r=0}$ and $\mathrm{\alpha = 0}$ in Eq. \eqref{reff} respectively. Figures \ref{comparecoherent} and \ref{coherency} show their individual plots.}

The phase resolution needed to tell apart different values of $\mathrm{r}$ is shown in Fig \ref{coherency}. Additionally, we also show in Fig. \ref{comparism} the interferometric phase resolution needed to measure the relative direction of squeezing with respect to the displacement. This shows our ability to distinguish two different coherent states as the amplitude of the state changes.

\subsection{QND measurement of Fock states using coherent states as reference}
 
The original formulation of mode invisibility \cite{marvy2013} considered  distinguishing between a Fock state containing $\mathrm{n}$ photons and another Fock state containing $\mathrm{n+m}$ photons.  The phase acquired by an atom crossing a cavity containing a Fock state of light is \cite{marvy2013}
\begin{align*}
\gamma(n) &=\! \operatorname{Re}\Bigg\{\!\!\!-\!\ii \operatorname{ln}\! \bigg[ 1\! - \!{\lambda^{2}} \bigg(\frac{(C_{+,\kappa}^{*} + C_{-,\kappa})}{k_{\kappa}L}n+\! \sum_{\beta}\!  \frac{C_{+,\beta}^{*}}{k_{\beta}L} \bigg) \bigg]\!\Bigg\}
\end{align*}

 In the non-relativistic case, if we assume $\mathrm{\gamma \ll 1}$, then this expression is given as
\begin{equation*}
e^{i\eta} \sim 1+\ii\eta= 1 - \! \lambda^{2} \bigg[n\frac{C_{-,\alpha}}{k_{\alpha}L}+\! \sum_{\beta\neq \alpha} \frac{C_{+, \beta}^{*}}{k_{\beta}L} +  (n+1)\frac{C_{+,\alpha}^{*}}{k_{\alpha}L} \bigg]
\end{equation*}
and so the phase $\mathrm{\gamma=\text{Re} (\eta)}$ will be given by
\begin{equation}\label{phase2}
\gamma\simeq - \text{Im}\left({\lambda^{2}} \bigg[n\frac{C_{-,\kappa}}{k_{\alpha}L}+\! \sum_{\beta\neq \kappa} \frac{C_{+, \beta}^{*}}{k_{\beta}L} +  (n+1)\frac{C_{+,\kappa}^{*}}{k_{\kappa}L} \bigg]\right).
\end{equation}

On the other hand,  the phase acquired by an atom crossing a cavity sustaining a coherent state of light is given in Eq. \eqref{roguephase}.  In the nonrelativistic case, if we assume $\mathrm{\gamma \ll 1}$, we also have
\begin{equation}\label{phazes}
\gamma\simeq - \text{Im}\left({\lambda^{2}} \bigg[\frac{C_{-,\kappa}}{k_{\kappa}L}|\alpha_{R}|^{2}+\! \sum_{\beta} \frac{C_{+, \beta}^{*}}{k_{\beta}L} +  |\alpha_{R}|^{2}\frac{C_{+,\kappa}^{*}}{k_{\kappa}L} \bigg]\right).
\end{equation}
   If we prepare an experimental setup as illustrated in Fig. \ref{scheme} to measure an unknown Fock state of light trapped in a cavity using a known coherent state as the reference, the difference between phases acquired during an interaction with this light field state is given in the nonrelativistic limit as $\Delta \gamma^{(\alpha)}_{n} =  \gamma(n)- \gamma(\alpha_{R})$, which therefore gives
\begin{align}\label{Fockcoherent}
\Delta \gamma^{(\alpha)}_{n}= \frac{\lambda^{2}}{k_{\kappa}L}\text{Im}\Big[\frac{(C_{+,\kappa}^{*}+C_{-,\kappa})}{k_{\kappa}L}\Big](n-|\alpha_{R}|^{2})
\end{align}

We can compare the phase resolution in this case (the interferometric phase difference between a Fock state of $\mathrm{n}$ photons and a Fock state of $\mathrm{n+m}$ photons). We defined the phase resolution of the interferometric experiment as   the difference in the observed interferometric phases between states with $\mathrm{n}$ and $\mathrm{n+m}$ photons.  Previously  \cite{marvy2013} we employed a known Fock state as a reference, yielding
\begin{align*}
\mathcal{R}_m(n) = \Delta\gamma (m+n) - \Delta\gamma(n),
\end{align*}
 where the respective interferometric phases are $\mathrm{\Delta\gamma (m+n)}$ and $\mathrm{\Delta\gamma (n)}$ for $\mathrm{m+n}$ and $\mathrm{n}$ photons. This phase resolution was found to respond linearly  for a small number of photons in the target state. However as the number of photons increases, the slope of the curve decreases logarithmically, worsening the resolution of the interferometric  experiment \cite{marvy2013}. This is not surprising: it is challenging to distinguish  a state with $10^{6}$ photons from one with $10^{6} + 1$  photons, unlike distinguishing between single-photon and two-photon states.
 
 As mentioned earlier, a significant challenge in the previously proposed   measurement scheme \cite{marvy2013}  is to prepare a Fock state of light in the reference cavity.  The discussion  above indicates that coherent states can be used as a reference for the interferometric phase determination of an unknown Fock state.
 
\begin{figure}[h!]
\begin{center}
\includegraphics[width=.40\textwidth]{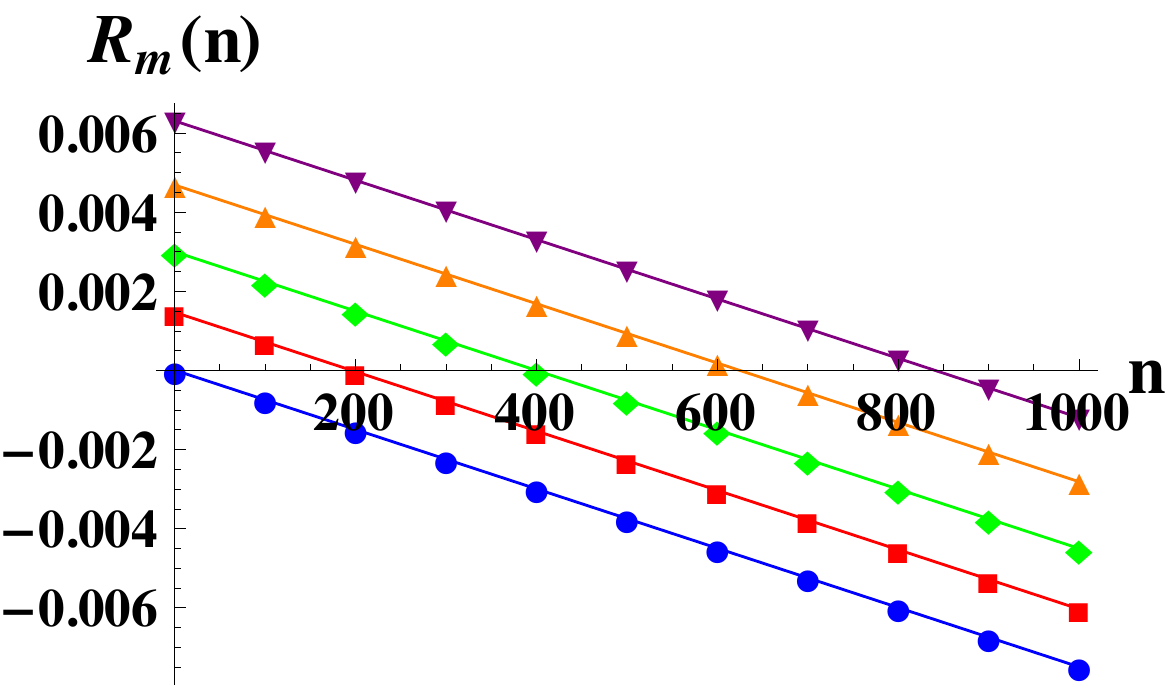}
\caption{(Color online): Phase resolution required to distinguish between a Fock state of unknown number $n$ of photons and another with $n+m$ photons, using a coherent state as a reference.  Here we show the plots for different values of $m = 5, 10, 15, 20, 25$); respectively.}
\label{compare}
\end{center}
\end{figure}

 In this new scenario,  the interferometric phase between the reference coherent state and the unknown Fock state that we want to identify is given by \eqref{Fockcoherent}. The phase resolution required to distinguish between a Fock state containing $\mathrm{n}$ photons and one containing $\mathrm{n+m}$ photons, in this new scheme where the reference is a known coherent state, is
\begin{align*}
\Delta_{m}^{(\alpha)}\gamma(n) = \Delta\gamma^{(\alpha)}_{n} -  \Delta\gamma^{(\alpha)}_{n+m}.
\end{align*}
In the nonrelativistic case, this becomes  
\begin{align*}
\Delta_{m}^{(\alpha)}\gamma(n) =-\frac{\lambda^{2}}{k_{\kappa}L}\text{Im}\Big[\frac{(C_{+,\kappa}^{*}+C_{-,\kappa})}{k_{\kappa}L}\Big]m,
\end{align*}
and we have the full expression (up to order $\lambda^2)$ given as
\begin{align*}
\Delta_{m}^{(\alpha)}\gamma(n) \!=\!\operatorname{Re} \left[\ii \operatorname{ln} \left( \frac{1\! -\! {\lambda^{2}} \bigg(\frac{nC_{+,\kappa}^{*}}{k_{\kappa}L}+\! \sum_{\beta}  \frac{C_{+,\beta}^{*}}{k_{\beta}L}\bigg) }{1\! -\! {\lambda^{2}} \bigg(\frac{(n+m)C_{+,\kappa}^{*}}{k_{\kappa}L}+\! \sum_{\beta}  \frac{C_{+,\beta}^{*}}{k_{\beta}L}\bigg)}\right)\right].
\end{align*}
The last equation yields an expression independent of the amplitude of the coherent state, one similar to that of a reference cavity sustaining a Fock state of known photon number $\mathrm{m}$. Figure \ref{compare} shows the phase resolution plotted against the unknown number of photons $\mathrm{n}$  in the Fock state that we want to probe. We see that the phase resolution is linearly dependent on the number of photons. Here we send the atom into the interferometer at speed $\mathrm{v = 1000 m/s}$ and a coupling strength $\mathrm{\lambda = 10^{-6}-10^{-4}}$. Given that the phase resolution required for an interferometric experiment is of the order of fractions of millirads, we see from the plot that we have more than enough phase resolution to distinguish between Fock states differing only in one photon using a coherent state reference.

\section{Stability of the Mode Invisibility}

As a final remark, notice that the mode-invisibility technique relies on the fact that the spatial symmetry of the even field modes effectively cancels the action that an atomic probe transversing the cavity at constant speed exerts on the field. For that, it is important that the speed of the atom crossing the cavity be kept constant. A legitimate question one may ask then is how stable the method is if we allow for variations in the speed, so that the atom spends more time in the first half of the cavity than it spends in the second.

\begin{figure}[h!]
\begin{center}
\includegraphics[width=.35\textwidth]{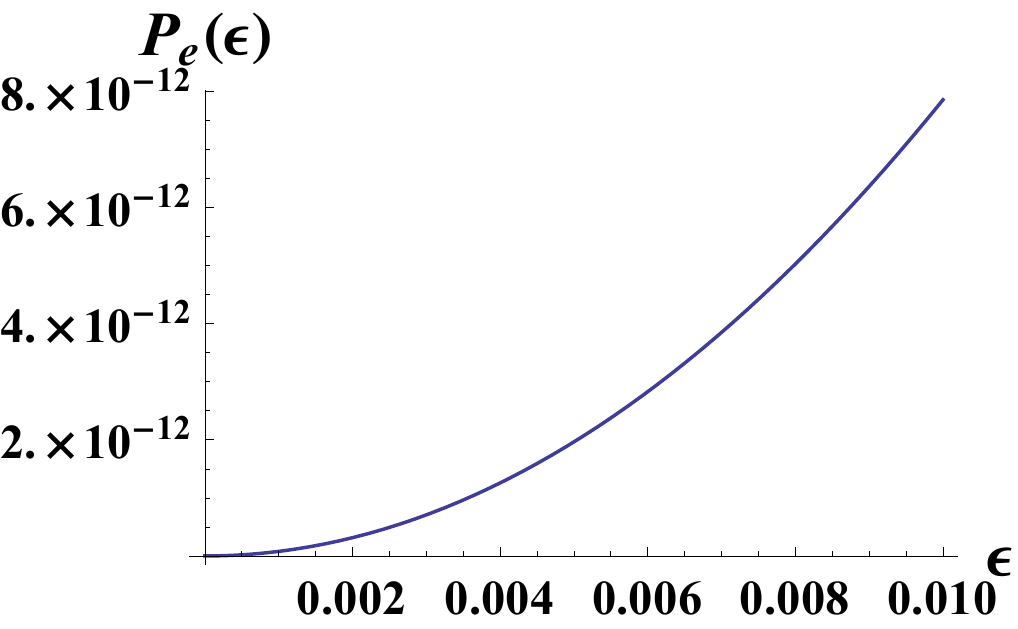}
\caption{(Color online): Excitation transition probability of an atom as a function of the switching parameter $\epsilon$. We can see that the value of $P_{e}(\epsilon)$ is small for the given values of $\epsilon$.}
\label{probability}
\end{center}
\end{figure}

A simple way  to assess this stability is to consider that the coupling strength is not uniform in time. For example we can have that the coupling strength of the atom in the first half of the cavity be larger than the coupling strength in the second half. With that purpose, we can monitor how the excitation transition probability is minimized by introducing a switching function $\chi(t) = (1-\epsilon t)$ to the interaction Hamiltonian
\begin{align*}
H_{I} = \lambda \chi(t) \mu(t) \phi[x(t)]
\end{align*}
This means that the leading order contribution to the amplitude of probability to change to a different state due to the passage of the atom \eqref{prob}, will be modified as
\begin{align}\label{probmod}
I_{\pm, \kappa} = \int_{0}^{T} \mathrm{d}t\, (1-\epsilon t)e^{\ii(\pm \Omega + \omega_{\kappa})t}\sin[k_{\kappa}x(t)].
\end{align}
And the probability that the systems evolve to a different state ---which under the weak adiabatic condition has to fulfill the condition that $P_e\ll1$--- is plotted as a function of $\epsilon$ in Fig. \ref{probability}.
This shows that the method is rather stable (quadratic in $\epsilon$) under small perturbations of the symmetry that allows the mode invisibility to work. The transition probability is approximately $P_{\ket{e}} \approx 10^{-14}$ for $\epsilon =10^{-3}$.

\section{Conclusions}

We have extended the   `mode-invisibility' technique proposed in \cite{marvy2013} to find information about the Wigner function of more general states of light in a nondestructive  measurement by means of atomic interferometry. Our result shows that the technique  can be safely employed to probe squeezed--coherent states. We showed that for realistic values of the physical parameters, it is possible to gain information about the amplitude of the squeezing parameter $r$, about the coherent state parameter $|\alpha|$ and, even more, about the relative phase difference $(2\theta - \phi)$, (the direction of the squeezing in phase space relative to the direction of displacement), all of it without significantly perturbing the quantum state of light probed. 
Since the method is a nondemolition measurement, we could employ successive measurements (which will not alter the state of light significantly) to characterize more than one parameter of the state of light.

Furthermore, how this `mode-invisibility' technique could  be used to characterize, in a nondemolition way, some features of the Wigner function of different states of light (photon number expectations, phase space distribution first and second moments, etc.) remains unexplored.

We also showed that we do not need to control and sustain known Fock states in a reference cavity in order to probe a target Fock state. Instead, an easily produced and sustained reference coherent state can be used,  yielding the same phase resolution as earlier obtained in \cite{marvy2013}. This eliminates the setbacks of having to control and maintain a reference Fock state to probe another one,   considerably simplifying a possible experimental implementation.

A natural direction to pursue for future work is to  see if our mode-invisibility approach can be used to extract more information about the Wigner function of the states of light. In particular, given the mean of the distribution, how sensitive is our approach
to the uncertainties in the quadratures?  These studies may lead to proposals for QND techniques that may help build less demanding techniques of state tomography of quantum light.

\section{Acknowledgments}
 The authors thank Achim Kempf for helpful and interesting comments about the mode invisibility technique.  This work was supported in part by the Natural Sciences and Engineering Research Council of Canada. E. M-M acknowledges support of the Banting Postdoctoral Fellowship Programme. 

\appendix
\begin{widetext}
 \section{Calculations for the atom-squeezed coherent field system}\label{firstappendix}
\subsection{Excitation transition probability}
Here we will estimate the excitation transition probability of the atom- squeezed coherent field system given by the expression $\langle e | \operatorname{Tr}_{F}[U^{(1)}\rho(0) U^{(1)\dagger}]| e \rangle$; with $\rho(0) = \ket{g}\langle g | \otimes \ket{\zeta,\alpha}_{\kappa}\langle \zeta,\alpha|_{\kappa}$, where the notation shows implicitly that all the other modes of the field are in the vacuum state. We will write $\rho(0)$ as $\rho$ for convenience.
\begin{align*}
\langle e|\operatorname{Tr}_{F}[U^{(1)}\rho U^{(1)\dagger}]|e\rangle =&\bigg[\sum_{\gamma} \sigma^{+}a^{\dagger}_{\gamma}I_{+,\gamma} + \sum_{\gamma}\sigma^{+}a_{\gamma}I_{-\gamma}^{*}\bigg] \bigg( \ket{g}\langle g | \otimes \ket{\zeta,\alpha}_{\kappa}\langle \zeta,\alpha|_{\kappa}\bigg) \bigg[\sum_{\beta}\sigma^{-}a_{\beta} I_{+,\beta}^{*} + \sum_{\beta}\sigma^{-}a^{\dagger}_{\beta}I_{-,\beta}\bigg]
\end{align*}
Rearranging these terms and taking note of the cyclic properties of the trace of products of operators, we have
\begin{align*}
\operatorname{Tr}_{F}[U^{(1)}\rho U^{(1)\dagger}] =&\bigg[ \operatorname{Tr}_{F}\big[\ket{e}\langle e|\sum_{\beta,\gamma}I_{+,\beta}^{*}I_{+,\gamma} \langle \zeta, \alpha|_{\kappa} a_{\beta}a^{\dagger}_{\gamma}\ket{\zeta,\alpha}_{\kappa} \big] +\operatorname{Tr}_{F}\big[\ket{e}\langle e|\sum_{\beta,\gamma}I_{-,\beta}I_{-,\gamma}^{*} \langle \zeta,\alpha|_{\kappa}a_{\beta}^{\dagger}a_{\gamma}\ket{\zeta,\alpha}_{\kappa} \big]\\
&+\operatorname{Tr}_{F}\big[\ket{e}\langle e|\sum_{\beta,\gamma}I_{-,\beta}I_{+,\gamma} \langle \zeta, \alpha|_{\kappa}a_{\beta}^{\dagger}a_{\gamma}^{\dagger}\ket{\zeta,\alpha}_{\kappa} \big]   + \operatorname{Tr}_{F}\big[\ket{e}\langle e|\sum_{\beta,\gamma}I_{+,\beta}^{*}I_{-,\gamma}^{*} \langle \zeta,\alpha|_{\kappa} a_{\beta}a_{\gamma}\ket{\zeta,\alpha}_{\kappa} \big] \Bigg]
\end{align*}
It is easy to check that, applying the mode invisibility technique that ensures the detector to probe the even relevant field modes, the last two terms vanish for non-relativistic speeds $v/c \ll 1$. Therefore we are left with the first two terms. Taking note of the canonical relationship $a_{\beta}a_{\gamma}^{\dagger} = \delta_{\gamma\beta} + a^{\dagger}_{\gamma}a_{\beta}$, we can rewrite the first two terms.
\begin{align}\label{finals}
\operatorname{Tr}_{F}[U^{(1)}\rho U^{(1)\dagger}] =&\bigg[ \operatorname{Tr}_{F}\big[\ket{e}\langle e|\sum_{\beta,\gamma}I_{+,\beta}^{*}I_{+,\gamma}\langle \zeta, \alpha|_{\kappa}\delta_{\gamma,\beta}\ket{\zeta,\alpha}_{\kappa} \big] +\operatorname{Tr}_{F}\big[\ket{e}\langle e|\sum_{\beta,\gamma} M_{\beta, \gamma} \langle \zeta, \alpha|_{\kappa}a_{\beta}^{\dagger}a_{\gamma}\ket{\zeta,\alpha}_{\kappa} \big]\Bigg]
\end{align}
where for notational convenience we have written $M_{\beta,\gamma} = I_{-,\beta} I_{-,\gamma}^{*} + I_{+,\beta}^{*}I_{+,\beta}$. We will evaluate the sum in the second term in this expression by writing the state $\ket{\zeta,\alpha}_{\kappa}  = S(\zeta)_{\kappa}D(\alpha)_{\kappa}\ket{0}$. Keeping the sum over one variable $\beta$ constant and splitting the sum over $\gamma$ in two terms ($\gamma = \kappa$ and $\gamma \neq \kappa$) yields
\begin{align*}
\sum_{\beta,\gamma}M_{\beta,\gamma} \langle \zeta, \alpha|_{\kappa}a_{\beta}^{\dagger}a_{\gamma}\ket{\zeta,\alpha}_{\kappa} =& \sum_{\beta,\gamma \neq \kappa}M_{\beta,\gamma} \langle 0|D_{\kappa}^{\dagger}(\alpha)S_{\kappa}^{\dagger}(\zeta)a_{\beta}^{\dagger}a_{\gamma}S_{\kappa}(\zeta)D_{\kappa}(\alpha)\ket{0} \\
&+ \sum_{\beta,\gamma = \kappa}M_{\beta,\gamma}  \langle 0|D_{\kappa}^{\dagger}(\alpha)S_{\kappa}^{\dagger}(\zeta)a_{\beta}^{\dagger}a_{\gamma}S_{\kappa}(\zeta)D_{\kappa}(\alpha)\ket{0}
\end{align*}
The first term on the right hand side vanishes since $a_{\gamma}S_{\kappa}(\zeta)D_{\kappa}(\alpha)\ket{0} =S_{\kappa}(\zeta)D_{\kappa}(\alpha) a_{\gamma}\ket{0} = 0 $  for $\gamma \neq \kappa$. Therefore
\begin{align*}
\sum_{\beta,\gamma}M_{\beta,\gamma}  \langle \zeta, \alpha|_{\kappa}a_{\beta}^{\dagger}a_{\gamma}\ket{\zeta,\alpha}_{\kappa} =&\sum_{\beta}M_{\beta,\kappa}  \langle 0|D_{\kappa}^{\dagger}(\alpha)S_{\kappa}^{\dagger}(\zeta)a_{\beta}^{\dagger}a_{\kappa}S_{\kappa}(\zeta)D_{\kappa}(\alpha)\ket{0}
\end{align*}
Similarly separating the sum over the variable $\beta$ into two parts, when $\beta=\kappa$ and when $\beta \neq \kappa$
\begin{align*}
\sum_{\beta,\gamma}M_{\beta,\gamma}  \langle \zeta, \alpha|_{\kappa}a_{\beta}^{\dagger}a_{\gamma}\ket{\zeta,\alpha}_{\kappa} =&\sum_{\beta\neq \kappa}M_{\beta,\kappa}  \langle 0|D_{\kappa}^{\dagger}(\alpha)S_{\kappa}^{\dagger}(\zeta)a_{\beta}^{\dagger}a_{\kappa}S_{\kappa}(\zeta)D_{\kappa}(\alpha)\ket{0} \\&+ \sum_{\beta= \kappa}M_{\beta,\kappa} \langle 0|D_{\kappa}^{\dagger}(\alpha)S_{\kappa}^{\dagger}(\zeta)a_{\beta}^{\dagger}a_{\kappa}S_{\kappa}(\zeta)D_{\kappa}(\alpha)\ket{0} 
\end{align*}
We will take each of these terms one by one. Also taking note that  for $\beta \neq \kappa$, we have $a_{\beta}^{\dagger}a_{\kappa}S_{\kappa}(\zeta)D_{\kappa}(\alpha)\ket{0} =a_{\kappa}S_{\kappa}(\zeta)D_{\kappa}(\alpha) a_{\beta}^{\dagger}\ket{0} = a_{\kappa}S_{\kappa}(\zeta)D_{\kappa}(\alpha)\ket{1}_{\beta}$;
\begin{align}\label{for}
\sum_{\beta\neq \kappa}M_{\beta,\kappa} \langle 0|D_{\kappa}^{\dagger}(\alpha)S_{\kappa}^{\dagger}(\zeta)a_{\beta}^{\dagger}a_{\kappa}S_{\kappa}(\zeta)D_{\kappa}(\alpha)\ket{0} = \sum_{\beta\neq \kappa}M_{\beta,\kappa} \langle 0|D_{\kappa}^{\dagger}(\alpha)S_{\kappa}^{\dagger}(\zeta)a_{\kappa}S_{\kappa}(\zeta)D_{\kappa}(\alpha)\ket{1}_{\beta};
\end{align}

\begin{align}\nonumber
\sum_{\beta\neq \kappa}M_{\beta,\kappa}  \langle 0|D_{\kappa}^{\dagger}(\alpha)S_{\kappa}^{\dagger}(\zeta)a_{\beta}^{\dagger}a_{\kappa}S_{\kappa}(\zeta)D_{\kappa}(\alpha)\ket{0} &= \sum_{\beta\neq \kappa}M_{\beta,\kappa}  \langle 0|D_{\kappa}^{\dagger}(\alpha)[a_{\kappa}\cosh(r) - a_{\kappa}^{\dagger}e^{\ii \phi}\sinh(r)]D_{\kappa}(\alpha)\ket{1}_{\beta}\\
&\!\!\!\!\!\!\!\!\!\!\!\!\!\!\!\!\!= \sum_{\beta\neq \kappa}M_{\beta,\kappa}\left[ \cosh(r)\langle 0|D_{\kappa}^{\dagger}(\alpha)a_{\kappa} D_{\kappa}(\alpha)\ket{1}_{\beta}-e^{\ii \phi} \sinh(r)\langle 0|D_{\kappa}^{\dagger}(\alpha)a_{\kappa} D_{\kappa}(\alpha)\ket{1}_{\beta}\right].
\end{align}

Substituting Eqs. \eqref{tone} and \eqref{ttwo}, we have
\begin{align*}
\sum_{\beta\neq \kappa}M_{\beta,\kappa} \langle 0|D_{\kappa}^{\dagger}(\alpha)S_{\kappa}^{\dagger}(\zeta)a_{\beta}^{\dagger}a_{\kappa}S_{\kappa}(\zeta)D_{\kappa}(\alpha)\ket{0} = & \sum_{\beta\neq \kappa}M_{\beta,\kappa}\alpha \cosh(r)\braket{0}{1}_{\beta}   -  
\sum_{\beta\neq \kappa}M_{\beta,\kappa}\alpha^{*} e^{\ii \phi}\sinh(r)\braket{0}{1}_{\beta} \\
& + \sum_{\beta\neq \kappa}M_{\beta,\kappa}\cosh(r)\langle 0|a_{\kappa}\ket{1}_{\beta}  -  \sum_{\beta\neq \kappa}M_{\beta,\kappa}\alpha^{*} e^{\ii \phi}\sinh(r)\langle 0|a_{\kappa}^{\dagger}\ket{1}_{\beta} 
\end{align*}
The terms on the right hand side give zero contribution. Therefore the term $\sum_{\beta\neq \kappa}M_{\beta,\kappa} \langle 0|D_{\kappa}^{\dagger}(\alpha)S_{\kappa}^{\dagger}(\zeta)a_{\beta}^{\dagger}a_{\kappa}S_{\kappa}(\zeta)D_{\kappa}(\alpha)\ket{0}$ in Eq. \eqref{for} has no contribution to the excitation transition probability. The next term we would evaluate is
\begin{align}\label{hedge}
 \sum_{\beta= \kappa}M_{\beta,\kappa} \langle 0|D_{\kappa}^{\dagger}(\alpha)S_{\kappa}^{\dagger}(\zeta)a_{\beta}^{\dagger}a_{\kappa}S_{\kappa}(\zeta)D_{\kappa}(\alpha)\ket{0} 
=M_{\kappa,\kappa} \langle 0|\underbrace{D_{\kappa}^{\dagger}(\alpha)S_{\kappa}^{\dagger}(\zeta)a_{\kappa}^{\dagger}S(\zeta)_{\kappa}S_{\kappa}^{\dagger}(\zeta)a_{\kappa}S_{\kappa}(\zeta)D_{\kappa}(\alpha)}_{H}\ket{0} 
\end{align}
where we have made use of the unitarity of the squeeze operator $S(\zeta)_{\kappa}S_{\kappa}^{\dagger}(\zeta) = 1$.  We will go ahead to evaluate $H$. First we note that from Eq. \eqref{transformation}, 
\begin{align*}
b^{\dagger}_{\kappa}b_{\kappa}& = a_{\kappa}^{\dagger}a_{\kappa} \cosh^{2}(r) - a_{\kappa}^{\dagger}a_{\kappa}^{\dagger} e^{\ii \phi}\cosh(r)\sinh(r)  - a_{\kappa}a_{\kappa}e^{-\ii \phi}\cosh(r)\sinh(r) + \sinh^{2}(r)a_{\kappa}a^{\dagger}_{\kappa}\\
H =  D^{\dagger}_{\kappa}(\alpha)b^{\dagger}_{\kappa}b_{\kappa}  D_{\kappa}(\alpha) &=  \alpha a_{\kappa}^{\dagger} \cosh^{2}(r)  + \alpha^{*}a_{\kappa}\cosh^{2}(r)  + a_{\kappa}^{\dagger}a_{\kappa} \cosh^{2}(r) 
- e^{\ii \phi}\cosh(r)\sinh(r)  a_{\kappa}^{\dagger}a_{\kappa}^{\dagger}  + \alpha^{*}\sinh^{2}(r)a_{\kappa}\\
&- 2\alpha^{*} e^{\ii \phi}\cosh(r)\sinh(r) a_{\kappa}^{\dagger}  -e^{-\ii \phi}\cosh(r)\sinh(r)a_{\kappa}a_{\kappa} - 2 \alpha e^{-\ii \phi}\cosh(r)\sinh(r) a_{\kappa} +\alpha \sinh^{2}(r) a^{\dagger}_{\kappa}  \\
& + \sinh^{2}(r)a_{\kappa}a^{\dagger}_{\kappa}+ |\alpha|^{2}\cosh^{2}(r) + |\alpha|^{2}\sinh^{2}(r) - (\alpha^{*})^{2}e^{\ii \phi}\cosh(r)\sinh(r) -  \alpha^{2}e^{-2\ii \phi}\cosh(r)\sinh(r)
\end{align*}
Substituting these terms back in Eq. \eqref{hedge}, the only non-zero terms are the terms in the last line of the equation above. We therefore have
\begin{align*}
H =  D^{\dagger}_{\kappa}(\alpha)b^{\dagger}_{\kappa}b_{\kappa}  D_{\kappa}(\alpha) &= \sinh^{2}(r)a_{\kappa}a^{\dagger}_{\kappa}+ |\alpha|^{2}\cosh^{2}(r) + |\alpha|^{2}\sinh^{2}(r) - (\alpha^{*})^{2}e^{\ii \phi}\cosh(r)\sinh(r) -  \alpha^{2}e^{-2\ii \phi}\cosh(r)\sinh(r)\\
=&  \sinh^{2}(r)a_{\kappa}a^{\dagger}_{\kappa}+ |\alpha|^{2}[\cosh^{2}(r) +\sinh^{2}(r)] - 2\operatorname{Re}[(\alpha^{*})^{2}e^{\ii \phi}]\cosh(r)\sinh(r) 
\end{align*}
Substituting this back in Eq. \eqref{hedge}, we have
\begin{align}
\nonumber\sum_{\beta= \kappa}M_{\beta,\kappa}\langle 0|D_{\kappa}^{\dagger}(\alpha)S_{\kappa}^{\dagger}(\zeta)a_{\beta}^{\dagger}a_{\kappa}S_{\kappa}(\zeta)D_{\kappa}(\alpha)\ket{0}=&M_{\kappa,\kappa} \langle 0|  \sinh^{2}(r)a_{\kappa}a^{\dagger}_{\kappa}+ |\alpha|^{2}[\cosh^{2}(r) +\sinh^{2}(r)] \\
&- 2\operatorname{Re}[(\alpha^{*})^{2}e^{\ii \phi}]\cosh(r)\sinh(r) \ket{0} 
\end{align}
Similarly substituting back in Eq. \eqref{finals}, we have
\begin{align*}
\operatorname{Tr}_{F}[U^{(1)}\rho U^{(1)\dagger}] &= \ket{e}\langle e|M_{\kappa,\kappa}\Bigg( \sinh^{2}(r)+ |\alpha|^{2}[\cosh^{2}(r) +\sinh^{2}(r)] - 2\operatorname{Re}[(\alpha^{*})^{2}e^{\ii \phi}]\cosh(r)\sinh(r)+ \sum_{\gamma}|I_{+,\gamma}|^{2} \Bigg)
\end{align*}

Therefore the probability that a joint atom--squeezed coherent system would get excited after an interaction time $T$ is given by the expression (up to order $\lambda^2$)
\begin{align*}
P_{\ket{e}} &=\Bigg[\frac{\lambda^{2}}{k_{\kappa}L} ( | I_{-,\kappa}|^{2} + | I_{+,\kappa}|^{2}) \big[\cosh^{2}(r) + \sinh^{2}(r)\big]|\alpha|^{2}
 - \frac{2\lambda^{2}}{k_{\kappa}L} (| I_{-,\kappa}|^{2} + | I_{+,\kappa}|^{2}) \sinh(r) \cosh(r) \operatorname{Re}\big[| \alpha|^{2} e^{ \ii(2\theta - \phi)} \big]\\
&\qquad 
+( |I_{-,\kappa}|^{2} + |I_{+,\kappa}|^{2})\frac{\lambda^{2}\sinh^{2}(r)}{k_{\kappa}L}  + \sum_{\gamma}\frac{\lambda^{2}|I_{+,\gamma}|^{2}}{k_{\gamma}L}\Bigg].
\end{align*}

\subsection{Estimating the phase factor}
To calculate the phase acquired by the atom while it flies through a cavity with a coherent light field sustained in it, we need to evaluate the expression (see Sec \ref{weaker})
\begin{align*}
\eta = &- \ii \operatorname{ln}\Bigg[ 1 - \langle \psi(0) | U^{(2)}(T)| \psi(0) \rangle\Bigg]
\end{align*}
We need to evaluate $|\psi^{(2)}(T)\rangle$; which is defined as $U^{(2)}(T)| \psi(0) \rangle$. Substituting Eq. \eqref{evol2}, the only surviving term will be
\begin{align}\label{u2}
U^{(2)}(T)| \psi(0) \rangle =& \ket{g} \otimes \sum_{\gamma \beta}\Big[c_{\gamma \beta}^{+} a_{\gamma}a_{\beta}^{\dagger} + c^{-}_{\gamma \beta} a_{\gamma}^{\dagger}a_{\beta} \Big] \ket{\zeta, \alpha}_{\kappa} = \ket{g} \otimes \sum_{\gamma \beta}\Big[ c^{+}_{\gamma, \beta}\delta_{\beta \gamma} + (c^{+}_{\gamma \beta} + c^{-}_{\gamma \beta} )a_{\beta}^{\dagger}a_{\gamma} \Big]\ket{\zeta, \alpha}_{\kappa}\end{align}
where 
\begin{align*}
c^{\pm}_{\gamma,\beta} = \int_{0}^{L/v}dt \int_{0}^{t}dt' e^{\ii(\omega_{\gamma} \pm \Omega)t}e^{\ii(\omega_{\beta} \pm \Omega)t'} \sin(k_{\gamma}vt)\sin(k_{\beta}vt')
\end{align*}
Multiplying Eq. \eqref{u2} from the left with $\bra{\psi(0)}=\langle \alpha,\zeta |_{\kappa} $, we have
\begin{align*}
\langle \psi(0) | U^{(2)}(T)| \psi(0) \rangle &=  \sum_{\gamma}c^{+}_{\gamma} +\sum_{\beta,\gamma} \langle \alpha,\zeta|_{\kappa}(c^{+}_{\gamma \beta} + c^{-}_{\gamma \beta})a_{\gamma}^{\dagger}a_{\beta} \ket{\zeta,\alpha}_{\kappa}\\
=&  \sum_{\gamma}c^{+}_{\gamma} +\sum_{\beta,\gamma}C_{\beta,\gamma} \langle 0|D^{\dagger}(\alpha)_{\kappa}S^{\dagger}(\zeta)_{\kappa}a_{\gamma}^{\dagger}a_{\beta}S(\zeta)(\zeta)D_{\kappa}(\alpha) \ket{0}
\end{align*}

This is equivalent to the lengthy steps we took in deriving the transition probability for the joint system where we defined $C_{\gamma, \beta} = (c^{+}_{\gamma \beta} + c^{-}_{\gamma \beta})$. Therefore  the phase acquired by a detector after it has crossed a cavity sustaining a squeezed coherent state is (up to order $\lambda^2$)
\begin{align}\nonumber
\eta(r,\alpha)& = -\ii \operatorname{ln}\Bigg[1 - \lambda^{2}\bigg( \frac{( C_{+,\kappa}^{*}+C_{-,\kappa})}{k_{\kappa}L} \sinh^{2}(r)  + \sum_{\gamma}\frac{C_{+,\gamma}^{*}}{k_{\gamma}L}+\frac{(C_{+,\kappa}^{* } + C_{-,\kappa})}{k_{\kappa}L}  (\cosh^{2}(r) + \sinh^{2}(r))|\alpha|^{2} \\
&\qquad \qquad \qquad \qquad -\frac{ 2}{k_{\kappa}L}( C_{+,\kappa}+C_{-,\kappa}) \sinh(r) \cosh(r) \operatorname{Re}\big[| \alpha|^{2} e^{ \ii(2\theta - \phi)} \big] \bigg)\Bigg]
\end{align}

 Note that in the Appendix of \cite{marvy2013}, there is a sign typo in the expression for the term $R$ in the equation right below Eq. (C4) in \cite{marvy2013}. The term $R$ has the opposite sign than the one it should. However; the nonrelativistic limit [expression (C7) in \cite{marvy2013}] is correct. 

\end{widetext}

\bibliography{references}

\end{document}